\pdfoutput=1 
\documentclass{aa}
\usepackage{graphicx}
\usepackage{txfonts}
\usepackage[flushleft]{threeparttable}
\usepackage{multirow}
\usepackage{multicol}
\usepackage{amsmath}
\usepackage{hyperref}
\begin{document}

   \title{Survey of Surveys I}
   \subtitle{The largest catalogue of radial velocities for the Galaxy}
   \author{M. Tsantaki\inst{\ref{oaa}} 
        \and 
        E. Pancino\inst{\ref{oaa}, \ref{asi}}
        \and 
        P. Marrese\inst{\ref{oar}, \ref{asi}}
        \and 
        S. Marinoni\inst{\ref{oar}, \ref{asi}}
        \and
        M. Rainer\inst{\ref{oaa}}
        \and
        N. Sanna\inst{\ref{oaa}}
        \and
        A. Turchi\inst{\ref{oaa}}
        \and
        S. Randich\inst{\ref{oaa}}
        \and
        C. Gallart\inst{\ref{iac}, \ref{laguna}}
        \and
        G. Battaglia\inst{\ref{iac}, \ref{laguna}} 
        \and
        T. Masseron\inst{\ref{iac}, \ref{laguna}}
        }

   \institute{INAF -- Osservatorio Astrofisico di Arcetri, Largo Enrico Fermi 5, 50125 Firenze, Italy\label{oaa} \\
              \email{maria.tsantaki@inaf.it}
         \and
    Space Science Data Center -- ASI, Via del Politecnico SNC, I-00133 Roma, Italy\label{asi}
         \and
    INAF -- Osservatorio Astronomico di Roma, Via Frascati 33, I-00078, Monte Porzio Catone (Roma), Italy\label{oar}
    \and
    Instituto de Astrofísica de Canarias, 38205 La Laguna, Tenerife, Spain\label{iac}
    \and
    Departamento de Astrofísica, Universidad de La Laguna, E-38200 La Laguna, Tenerife, Spain\label{laguna}
}

   \date{Received June 04, 2021; accepted October 04, 2021}

  \abstract
   {In the present-day panorama of large spectroscopic surveys, the amount, diversity, and complexity of the available data continuously increase. The overarching goal of studying the formation and evolution of our Galaxy is hampered by the heterogeneity of instruments, selection functions, analysis methods, and measured quantities. 
   }
   {We present a comprehensive catalogue, the \textit{Survey of Surveys} (SoS), built by homogeneously merging the radial velocity (RV) determinations of the largest ground-based spectroscopic surveys to date, such as APOGEE, GALAH, {\em Gaia}-ESO, RAVE, and LAMOST, using {\em Gaia} as reference. This pilot study serves to prove the concept and to test the methodology that we plan to apply in the future to the stellar parameters and abundance ratios as well.}
   {We have devised a multi-staged procedure that includes: \textit{i}) the cross match between {\em Gaia} and the spectroscopic surveys using the official {\em Gaia} cross-match algorithm, \textit{ii}) the normalization of uncertainties using repeated measurements or the three-cornered hat method, \textit{iii}) the cross calibration of the RVs as a function of the main parameters they depend on (magnitude, effective temperature, surface gravity, metallicity, and signal-to-noise ratio) to remove trends and zero point offsets, and \textit{iv}) the comparison with external high-resolution samples, such as the {\em Gaia} RV standards and the Geneva-Copenhagen survey, to validate the homogenization procedure and to calibrate the RV zero-point of the SoS catalogue.}
   {We provide the largest homogenized RV catalogue to date, containing almost 11 million stars, of which about half come exclusively from {\em Gaia} and half in combination with the ground-based surveys. We estimate the accuracy of the RV zero-point to be about 0.16-0.31\,km\,s$^{-1}$ and the RV precision to be in the range 0.05-1.50\,km\,s$^{-1}$ depending on the type of star and on its survey provenance. We validate the SoS RVs with open clusters from a high resolution homogeneous samples and provide the systemic velocity of 55 individual open clusters. Additionally, we provide median RVs for 532 clusters recently discovered by {\em Gaia} data.}
   {The SoS is publicly available, ready to be applied to various research projects, such as the study of star clusters, Galactic archaeology, stellar streams, or the characterization of planet-hosting stars, to name a few. We also plan to include survey updates and more data sources in future versions of the SoS.}
   \maketitle

\section{Introduction}

The last decades, Galactic astronomy has seen a revolution mainly thanks to the large data sets from various surveys monitoring our Galaxy. In particular, there has been an exponential increase in spectroscopic data where sample sizes have grown from hundreds of stars to several hundred thousands observed principally by multi-object spectrographs. The dedicated Galactic spectroscopic surveys have used this technology delivering a plethora of interesting results on the structure and evolution of the Milky Way, mainly from: the Radial Velocity Experiment (RAVE, \citealt{Steinmetz2006}), the Large Sky Area Multi-Object Fiber Spectroscopic Telescope (LAMOST, \citealt{Zhao2012}), the {\em Gaia}-ESO Survey (GES, \citealt{Gilmore2012, Randich2013}), the GALactic Archaeology with HERMES (GALAH, \citealt{desilva2015}), the Apache Point Observatory Galactic Evolution Experiment (APOGEE, \citealt{Majewski2017}), and others. Among the surveys, the {\em Gaia} space mission \citep{Gaia2016} holds a unique place because it has provided high precision astrometric, photometric, and spectroscopic data for an unprecedented number of sources, for 1.3 billion, 1.6 billion and 7.2 million sources, respectively.

In order to fully exploit the wealth of data delivered from the above surveys, such as radial velocities (RVs), stellar atmospheric parameters, and chemical abundances, it is essential to understand in detail their properties, as well as their capabilities and limitations. Moreover, each survey focuses on different parts of the Galaxy or region of the parameter space, and uses different instruments and data analysis techniques, each having different strengths and weaknesses. A homogenization process is therefore necessary when combining data sets from various sources to reveal the subtle physical phenomena in our Galaxy, for instance the uniform chemical and kinematic properties of star clusters. As a standard procedure, each survey team compares their output parameters with benchmark samples to understand systematic effects, assess random errors, and evaluate different analysis methods for different resolution regimes, also highlighted in other works \citep[e.g.][]{Lee2015, Xiang2015, Deepak2018, Anguiano2018}. However, there are still only a few studies employing homogenization techniques to merge different literature samples in comprehensive catalogues, and they mostly focus on high-resolution spectroscopy  \citep[e.g.][]{Hinkel2014, Soubiran2016}.

In this proof-of-concept study, we aim at combining radial velocity measurements from the six large surveys mentioned above, merging them in a homogeneous catalogue, the \textit{Survey of Surveys} (SoS), which will be a valuable tool to the community for data mining studies in our Galaxy and its satellites. In a future study, we will also add atmospheric parameters (effective temperature, $T_{\rm eff}$, surface gravity, $\log g$, metallicity, $[M/H]$) and chemical abundances to the SoS. Precise and accurate RVs play an important role in understanding the structure of the Galaxy \citep[e.g.][]{Binney2000}, in detecting substructures in the halo and disc and thus unraveling the merger history of the Galaxy \citep[e.g.][]{Helmi2020}, the kinematics and membership of star clusters \citep[e.g.][]{Kharchenko2013}, the orbits of globular clusters \citep[e.g.][]{Balbinot2018}, or the determination of orbital parameters and properties of binaries \citep[e.g.][]{Helminiak2017} and planetary systems \citep[e.g.][]{Trifonov2020}. 

To achieve our goal, we will first place all surveys on a common (albeit perhaps arbitrary) reference system, defined as the RV zero point (ZP), and remove from each survey any trends with parameters the RVs depend, such as magnitude, $T_{\rm eff}$, $\log g$, iron metallicity ($[Fe/H]$)\footnote{Usually the iron abundance is used as a proxy for the overall metallicity of a star.}, or signal-to-noise ratio (S/N). Since no survey is completely free from bias, a comparison of the RVs from each survey with each other will reveal its intrinsic biases. Using {\em Gaia} as a reference system in this work is advantageous because it is the only survey with a significant number of stars in common with each of the ground based thanks to its sheer size and all-sky coverage (see Fig.~\ref{surveys_molleview}). 

In  Sect.~\ref{datasources}, we describe the data used in this work and in Sect.~\ref{xm} the cross match analysis of the surveys with {\em Gaia}. In Sect.~\ref{errors}, we explore the treatment of the RV errors using the information from the repeated measurements and the three-cornered hat method. In Sect.~\ref{rvcalibration}, we perform the internal RV homogenization process. We present our unified catalogue in Sect.~\ref{soscatalog} as well as the comparisons with external samples to validate the absolute calibration system. Finally, in Sect.~\ref{science}, we present the science validation of our results with open clusters.

\section{Data sources}\label{datasources}

\begin{table}
\caption{Number (N) of objects included per survey, N of unique star entries, N of stars in common with {\em Gaia} DR2, and R is the resolution. The N of unique stars is derived after the analysis in Sect.~\ref{xm} to identify duplicated sources and is different from the ones presented in the respectively survey releases.}
\centering
\scalebox{0.88}{
\begin{tabular}{l r r r c}
\hline \hline
Survey & N$_{stars}$ & N$_{unique\,stars}$ & N$_{common\,Gaia}$ & R \\
\hline
   APOGEE DR16 & 473\,306    & 437\,303    & 467\,155    & 22\,500 \\
   GALAH  DR2  & 342\,682    & 342\,682    & 342\,088    & 28\,000 \\
   GES    DR3  & 25\,533     & 25\,502     & 25\,440     & 16-45\,000\\
   LAMOST DR5  & 5\,348\,712 & 4\,150\,974 & 3\,943\,355 & 1\,800 \\
   RAVE   DR6  & 518\,387    & 451\,783    & 517\,095    & 7\,500 \\
   {\em Gaia} DR2 & 7\,224\,632 & 7\,224\,632 & -     & 11\,500 \\
\hline
\end{tabular}}
\label{table:0}
\end{table}

\begin{figure*}
\centering
\includegraphics[width=8.5cm, clip]{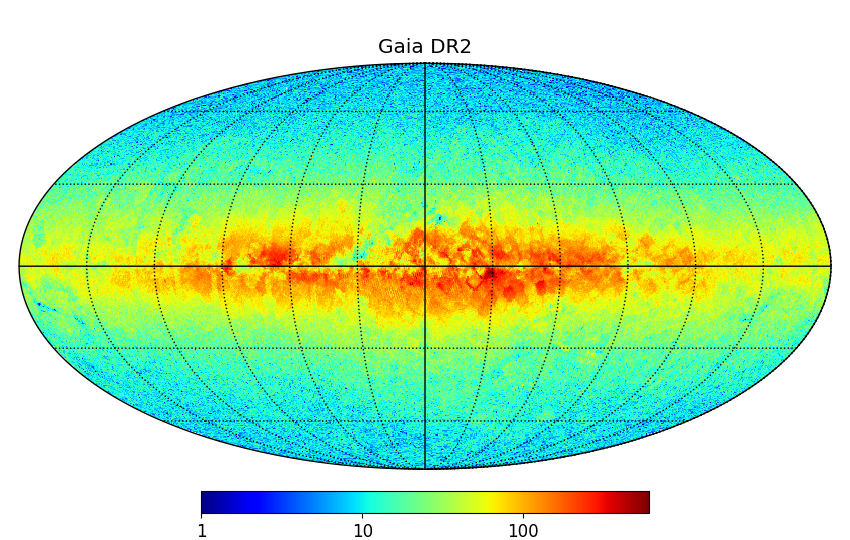} 
\includegraphics[width=8.5cm, clip]{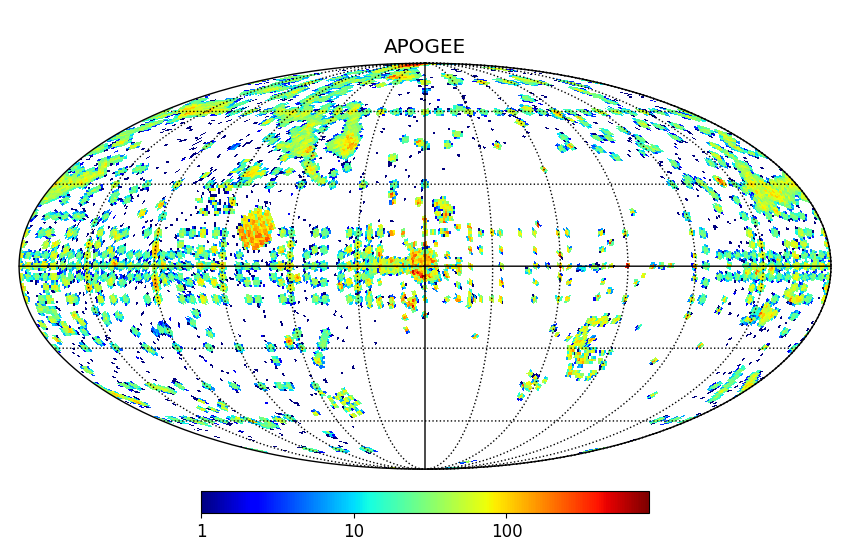} \\
\includegraphics[width=8.5cm, clip]{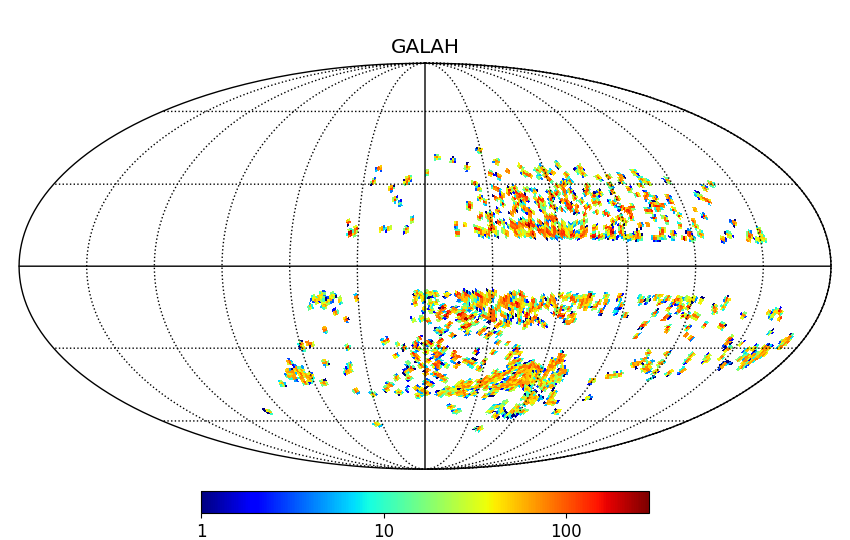} 
\includegraphics[width=8.5cm, clip]{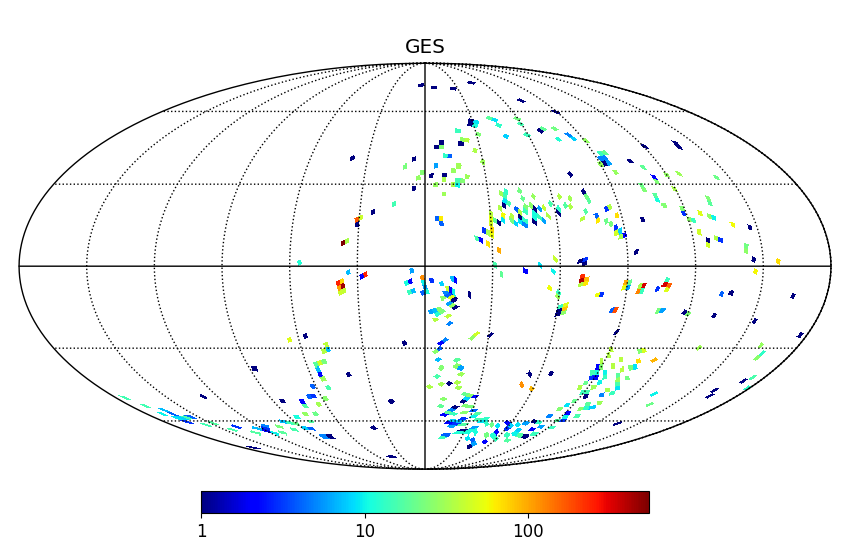}\\
\includegraphics[width=8.5cm, clip]{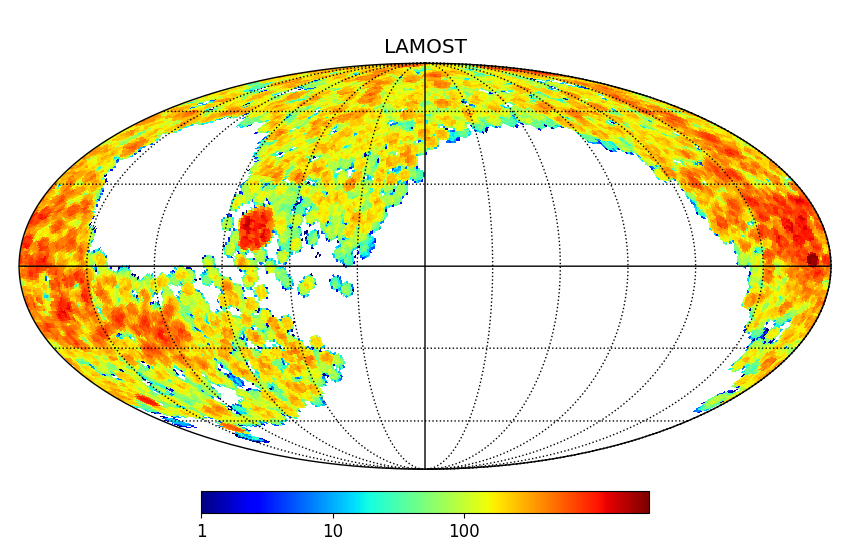} 
\includegraphics[width=8.5cm, clip]{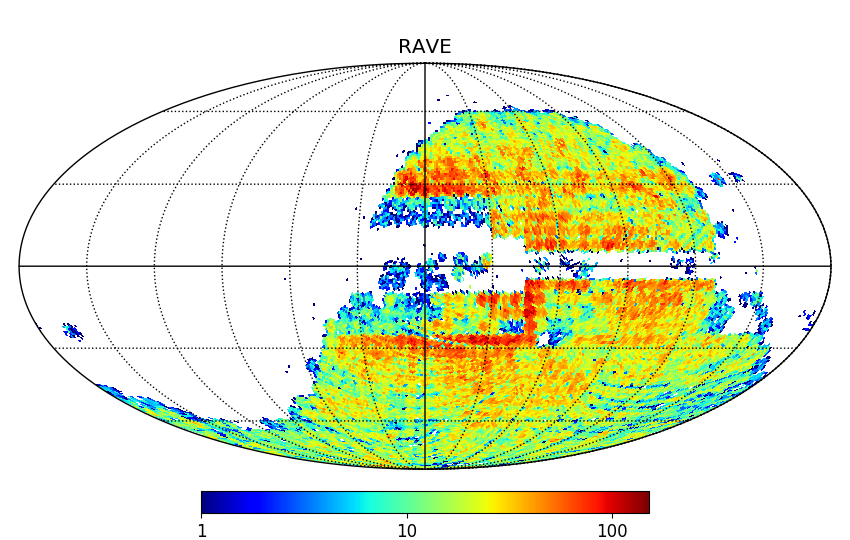} \\
\caption{Surface density distribution in a Mollweide projection of the galactic coordinates of the six surveys used in this work, obtained using a HEALPix (Hierarchical Equal Area isoLatitude Pixelization) tessellation with different resolutions. The color scales of the maps are in logarithmic scale and are different for each survey. }
\label{surveys_molleview}
\end{figure*}

\begin{figure}
\centering
\includegraphics[width=9.0cm, clip]{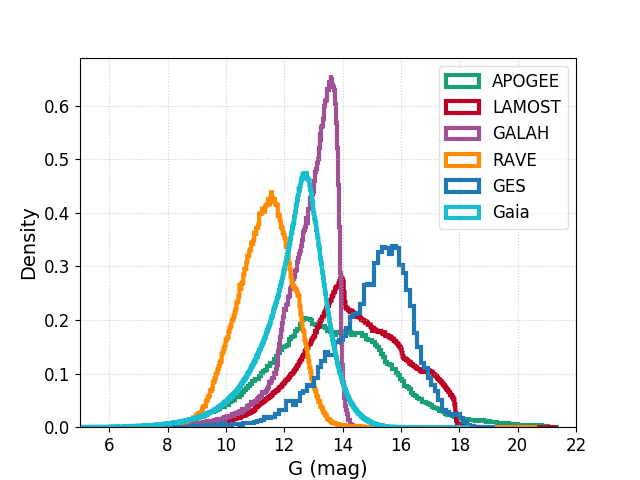} 
\caption{G magnitude distribution of the surveys in this work for stars in common with {\em Gaia}.}
\label{surveys_mag}
\end{figure}

\begin{figure*}
\hspace{-1.0cm}
\includegraphics[width=20cm, clip]{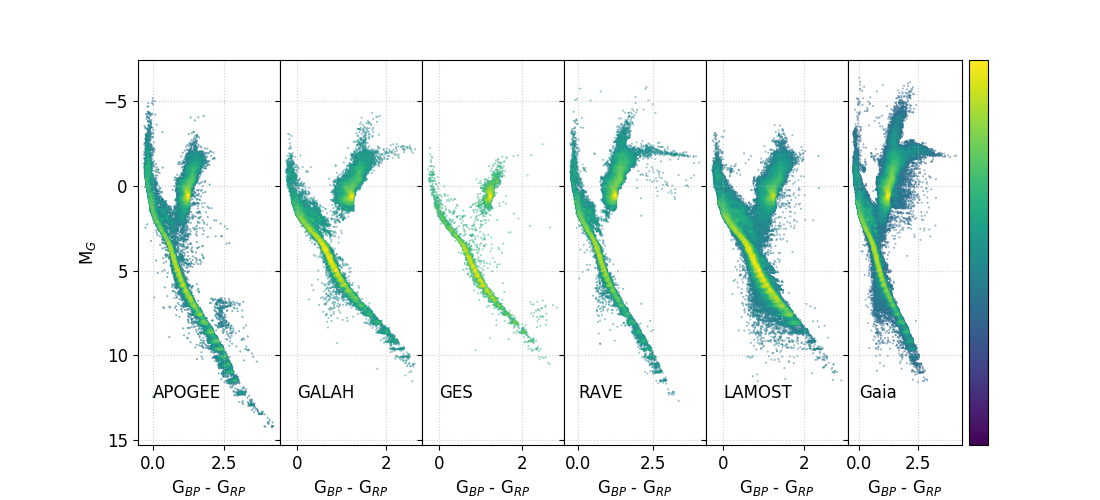} 
\caption{HR diagrams of the surveys used in this work using {\em Gaia} photometry and parallaxes, color coded to the stellar density in log scale.}
\label{cmd_surveys}
\end{figure*}

The core of the SoS is built on five ground-based spectroscopic surveys and {\em Gaia}. Each survey targets different stellar types but all of them have a common goal: to map the kinematics and chemistry of stars across the Galaxy to reconstruct its present-day structure, and past evolution history. In SoS\,I, we use the data releases (DR) of the surveys presented in Table~\ref{table:0} with their corresponding number of stars. 
Figures~\ref{surveys_molleview} and \ref{surveys_mag} show the distributions in surface density and magnitude of the considered data sets. In Fig.~\ref{cmd_surveys}, we present the Hertzsprung-Russell diagram (HRD) of the six surveys used in this work for stars with {\em Gaia} photometry available to showcase the stellar populations each survey targets. These plots are made after using the same filters as for the construction of the HRD for {\em Gaia} DR2 by \cite{Gaia2018} based on Eqs.~\ref{gaiafilter} (see below) and we have included the extinction and reddening magnitudes taken from {\em Gaia} DR2 as well.

\subsection{{\em Gaia} DR2}


The {\em Gaia} DR2 \citep{Brown2018} apart from the astrometric parameters (e.g. positions, parallaxes) has provided median radial velocities for 7.2 million stars, in the range of effective temperatures between 3550-6900\,K. The spectra have been collected by the Radial Velocity Spectrometer (RVS) on board, with resolution R\,(=\,$\lambda$/$\Delta\lambda$)\,$\simeq$\,11\,500 over the wavelength range 8450-8720\,\AA{} centred on the Calcium triplet. Each observed star is visited many times and the RVs are measured for each transit, as derived by a series of modules that compare the three CCD spectra corresponding to each field of view transit with a template of synthetic spectra \citep{Sartoretti2018}. The DR2 contains the median value from the multiple visits. The properties and validation of the {\em Gaia} RVs are described in detail in \cite{Katz2019}. The authors indicate that the {\em Gaia} radial velocity differences with respect to the ground-based surveys do not exceed 0.25-0.30\,km\,s$^{-1}$ and that the {\em Gaia} RVs show a positive trend as a function of magnitude (see Sect.~\ref{rvcalibration}). The overall precision estimated from the {\em Gaia} RV uncertainties is 1.05\,km\,s$^{-1}$ and mainly depends on the effective temperature and magnitude. 
The targets with available RVs cover the 4-13 G magnitude range (see Fig.~\ref{surveys_mag}). The {\em Gaia} DR3 will include updated radial velocities with atmospheric parameters to be published in 2022. Throughout this paper we use the {\em Gaia} DR2 source ID which should be treated independently from the DR3 even if the changes between the IDs in the catalogues are at the 2-3\% of stars \citep{Torra2020}.

We apply quality filtering criteria used for the RV calibrations in Sect.~\ref{rvcalibration} based on {\em Gaia} photometry suggested by \cite{Evans2018} and \cite{Arenou2018} to mitigate astrometric calibration problems and also contamination from double stars:
\begin{align}\label{gaiafilter}
1.0 + 0.015 (G_{BP} - G_{RP})^{2} < E < 1.3 + 0.06\,(G_{BP} - G_{RP})^{2}   \nonumber \\
u < 1.2\, \textrm{max}(1,  e^{-0.2(G - 19.5)}) \\
u = \sqrt{\chi^{2}/\nu+5}  \nonumber 
\end{align}
\\*
where G$_{BP}$ is the magnitude from the {\em Gaia} Blue Photometer (BP), G$_{RP}$ the magnitude from the {\em Gaia} Red Photometer (RP), E the \texttt{phot\_bp\_rp\_excess\_factor}\footnote{\texttt{phot\_bp\_rp\_excess\_factor} is the ratio of the sum of G$_{BP}$ and G$_{RP}$ fluxes over the G integrated flux.}, $\chi^{2}$ is the astrometric goodness-of-fit and, $\nu$ the number of good observations. 

We further flagged {\em Gaia} sources with potentially spurious RV estimations as indicated by \cite{Boubert2019} due to possible contamination by nearby objects\footnote{See also: https://www.cosmos.esa.int/web/gaia/dr2-known-issues} (see Sect.~\ref{soscatalog} for flags). The latter stars are flagged with the worse quality flag and comprise $\sim$1\%. {\em Gaia} DR2 does not contain duplicate sources. 

\subsection{APOGEE DR16}

The APOGEE and its successor APOGEE-2, is a high-resolution (R\,$\simeq$\,22\,500), high signal-to-noise (S/N\,$>$\,100 per half resolution element) spectroscopic survey using the 2.5\,m Sloan telescope in the northern hemisphere and the du Pont telescope at Las Campanas Observatory in the southern hemisphere. The survey operates in the near-infrared H-band, targeting mainly red giants.

In this work, we use the APOGEE DR16 which is the first release to include data from APOGEE-2, and therefore data from across the entire sky \citep{Ahumada2020}. The spectroscopic analysis of the 473\,307 spectra from 437\,303 unique stars observed in different telescope configurations is described in \cite{Jonsson2020}. APOGEE provides RVs calculated mainly in two ways: \textit{i}) the RVs were determined by cross-correlation iteratively of each visit spectrum against the combined observed spectrum since most stars are observed multiple times, and \textit{ii}) by cross-correlation against a best-matching synthetic template. The final RVs are selected as the ones to provide the smallest scatter in the individual RVs from either of the above methods (\texttt{VHELIO\_AVG} in the DR16). 

We applied the following filters to the APOGEE data to exclude stars from the calibration routines in Sect.~\ref{rvcalibration} but they are later calibrated given different quality flags. Our criteria exclude stars which exhibit significant differences in the RVs from the synthetic template and from those from the combined template (\texttt{STARFLAG} $\neq$ \texttt{SUSPECT\_RV\_COMBINATION}) for which we assign the lowest RV quality flag. We also exclude spectra with low S/N ($<$5) and erroneous stellar parameters (\texttt{STARFLAG} $\neq$ \texttt{LOW\_SNR}, \texttt{ASPCAP} $\neq$ \texttt{STAR\_BAD}) assigning them with an intermediate RV quality flag. After this filtering, we retained about 88\% of the measurements with the highest RV quality flag. 

\subsection{GALAH DR2}

The GALAH survey uses the High Efficiency and Resolution Multi-Element Spectrograph \citep{Sheinis2015} at the Anglo-Australian Telescope in high resolution (R\,$\simeq$\,28\,000) with four discrete optical wavelength channels: 4713-4903\,\AA{}, 5648-5873\,\AA{}, 6478-6737\,\AA{}, and 7585-7887\,\AA{}. The GALAH survey is an on-going spectroscopic survey to observe one million stars in the V magnitude range 12-14\,mag across the Southern sky. The data products of DR2 include RVs, stellar atmospheric parameters and chemical abundances for 32 different elements \citep{Buder2018}.

The RVs of GALAH DR2 are derived following the methodology of \cite{Zwitter2018} for 342\,682 stars. For the calculation of the RVs, stacked median observed spectra were used from multiple visits to build a reference library already shifted to rest wavelength. Then, for each star, its RV is obtained by comparison with the reference library via cross-correlation which is further improved by comparing with synthetic spectra from models which account for 3-dimensional convective motions in the stellar atmospheres. 
The GALAH DR2 does not contain any duplicate entries. The best quality stellar parameters are derived from their spectral analysis pipeline, CANNON, \citep{Buder2018} (\texttt{flag\_cannon}\,$=$\,0) which we used for the RV calibrations in Sect.~\ref{rvcalibration}, retaining 78\% of the stars. We note that the cleaner sample is used for calibration purposed but we provide calibrated RVs for the whole sample with a lower RV quality flag (see Sect~\ref{soscatalog}). The typical accuracy of the GALAH RVs is estimated at 0.1\,km\,s$^{-1}$ \citep{Zwitter2018}. 

We note that the third release of GALAH has just been published \citep{Buder2021} to include observations for the K2 and TESS follow-up programs as well as other ancillary observations, increasing the sample by 30\%. 

\subsection{{\em Gaia}-ESO DR3}

The {\em Gaia}-ESO survey is a large public survey \citep{Gilmore2012, Randich2013} carried out at the ESO Very Large Telescope (UT-2 Kueyen) with the FLAMES multi-object instrument \citep{Pasquini2002}. The survey has obtained high-quality spectra with the GIRAFFE spectrograph in different wavelength ranges depending on the spectrograph setting used (R\,$\simeq$\,16\,000-25\,000). The brighter stars however, have been observed with UVES (R\,$\simeq$\,45\,000) and correspond to about 10\% of the total sample. The GES targets cover a wide range of properties, from dwarfs to giants, from O to M stars focusing on relatively faint stars (mainly V\,$>$\,16\,mag), for which {\em Gaia} will not be able to provide accurate RVs and abundances.

The RVs from GIRAFFE spectra are derived based on cross-correlation with a grid of synthetic spectra to obtain an initial RV estimate and then by direct spectral fitting with a polynomial \citep{Koposov2011}. The RVs from UVES on the other hand, are derived via a standard cross-correlation method with a grid of synthetic template spectra at a range of temperatures, metallicities and gravities \citep{Sacco2014}. 

The best RV precision reached for the majority of the GES spectra from GIRAFFE spectrograph is 0.22–0.26\,km\,s$^{-1}$ for stars with low rotational broadening and large S/N and dependents on instrumental configuration in a study of repeated RV measurements of the same stars \citep{Jackson2015}. The typical RV error from the UVES spectra is 0.40\,km\,s$^{-1}$ \citep{Sacco2014}. The number of stars with available RVs in GES DR3 is 25\,533 and we did not apply any further quality filters for this survey. We obtained the GES DR3 from the public ESO archive\footnote{GES DR3: https://www.gaia-eso.eu/data-products/public-data-releases/gaia-eso-survey-data-release-3}.

\subsection{RAVE DR6}

We use the final data release of RAVE \citep{Steinmetz2020} which is magnitude-limited (9\,$<$\,I\,$<$\,12) and located in the southern hemisphere. The RAVE was set to use the 6dF multi-object spectrograph at the UK Schmidt Telescope from 2003 to 2013. The medium-resolution spectra (R\,$\simeq$\,7\,500) cover the Calcium triplet region (8410-8795\,\AA{}), a very similar wavelength range as the {\em Gaia} RVS.

The RAVE DR6 consists of, among other products, measurements of radial velocities, stellar atmospheric parameters, chemical abundances, and cross matches with other relevant catalogues for 518\,387 observations of 451\,783 stars \citep{Steinmetz2020,  Steinmetz2020b}. The RVs are derived with the pipeline SPARV which matches the observations to a grid of synthetic spectra with a standard cross-correlation algorithm in a two step process \citep[see details in][]{Zwitter2008}. 
The internal RV error distribution peaks at around 1.0\,km\,s$^{-1}$.

\cite{Steinmetz2020} suggest a set of quality criteria to select a clean sample of radial velocities, namely selecting stars based on \textit{i}) the zero-point correction applied to radial velocity ($|$\texttt{correctionRV}$|$\,$<$\,10\,km\,s$^{-1}$), \textit{ii}) the RV error (\texttt{hrv\_error\_sparv}\,$<$\,8\,km\,s$^{-1}$), and \textit{iii}) the Tonry–Davis correlation coefficient (\texttt{correlationCoeff}\,$>$\,10). These selection criteria define the core sample of the survey as indicated in \cite{Steinmetz2020}. The RV measurements after this filtering reach 69\% which are set with the lowest RV quality flag. 

\subsection{LAMOST DR5}

The LAMOST is a national scientific research facility operated by the Chinese Academy of Sciences in low-resolution (R\,$\simeq$\,1\,800) in the optical wavelength range (3650-9000\,\AA{}) and covering the northern hemisphere. The LAMOST Experiment for Galactic Understanding and Exploration (LEGUE) is an ongoing Galactic survey with a current sample of more than five million stellar spectra \citep{Deng2012} and is the largest survey considered here besides {\em Gaia}. LAMOST DR5 includes around 9 million objects but 5\,348\,712 of them consist of stars with published parameters, including 85\,845 A-type stars, 1\,694\,182 F-type stars, 2\,739\,467 G-type stars and 829\,218 K-type stars\footnote{LAMOST DR5: http://dr5.lamost.org/}.

The LAMOST pipeline \citep{Luo2015} measures the RVs using the cross-correlation method. The pipeline recognizes the stellar spectral classes and simultaneously determines the RVs from the best-fit correlation function between the observed spectra and the template. The RV error distribution of LAMOST peaks around 5\,km\,s$^{-1}$.

We adopted for LAMOST the quality cuts recommended by \cite{Luo2015}. In particular, we select spectra with S/N$>$15 where according to the survey lie the reliable parameters and exclude stars with negative RV errors. Hence, 8\% of the total sample which do not satisfy the above criteria have the lowest RV quality flag. 

\section{The cross match with Gaia}\label{xm}

\begin{table*}
\caption{The known duplicates are the stars with the same identification name provided from each survey. The possible duplicates are defined from the XM process. True duplicates are confirmed to be below the cut-off with 3 different criteria (q\,=\,99\%, q\,=\,97.5\%, 3\,$\sigma$). These stars are now treated as duplicated sources assigned with a new identification.}
\centering
\begin{tabular}{l r r r r r}
\hline \hline
\multirow{2}{*}{Survey} & Known  & Possible & true duplicates &  true duplicates &  true duplicates  \\
  & duplicates & duplicates & q\,=\,99\% &  q\,=\,97.5\% & 3\,$\sigma$ \\
\hline
   APOGEE & 64\,398     & 476      & 464 & 464      & 381    \\
   GALAH  & 0           & 0        & 0        & 0        & 0      \\
   GES    & 0           & 68       & 64       & 62       & 58     \\
   LAMOST & 1\,751\,152 & 529\,059 & 520\,310 & 518\,142 & 469\,008 \\
   RAVE   & 113\,964    & 14\,461  & 14\,371  & 14\,367  & 14\,105  \\
\hline
\end{tabular}
\label{table:1}
\end{table*}

\begin{figure}
\centering
\includegraphics[width=9.0cm, clip]{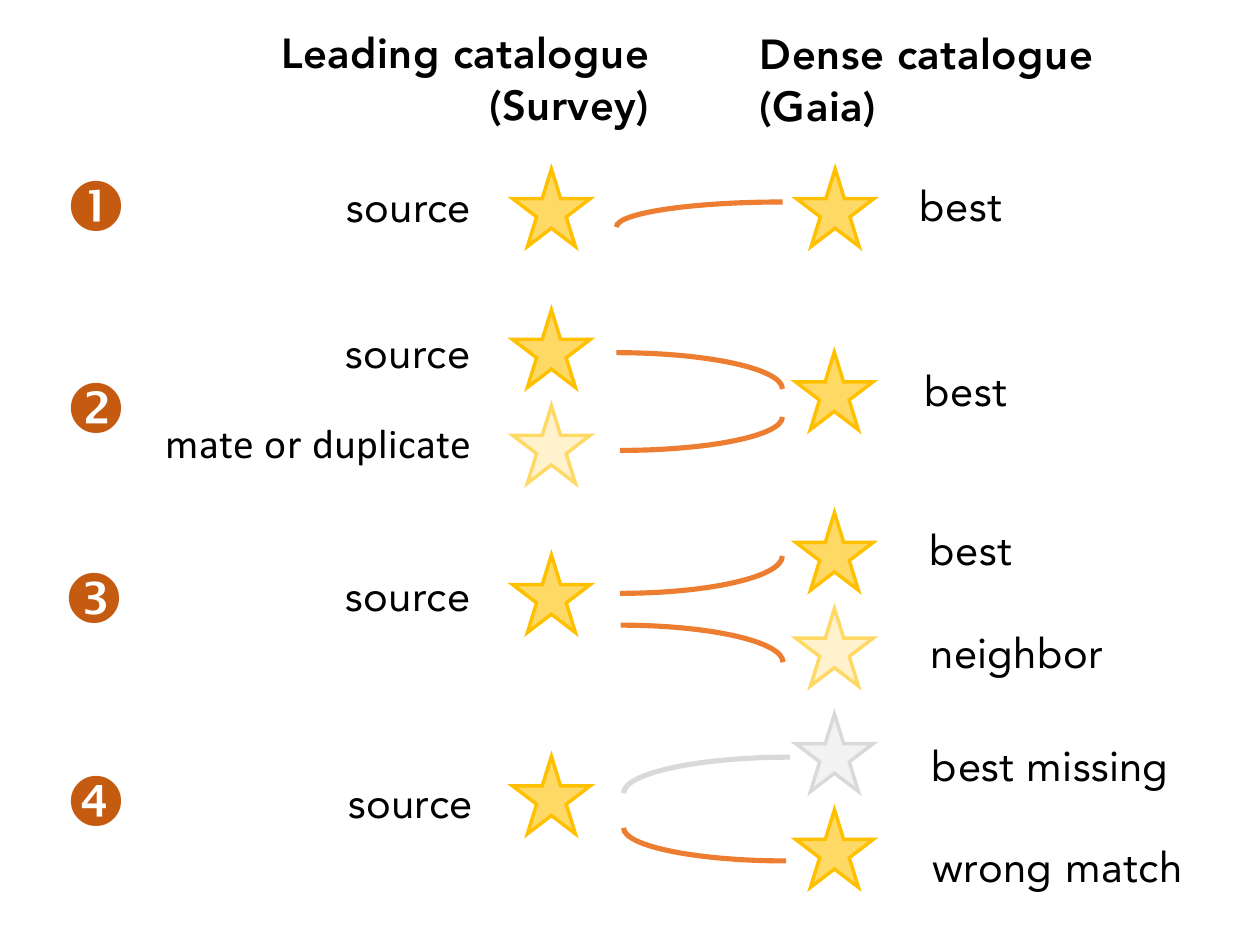} 
\caption{A sketch of four possible scenarios for the XM algorithm. Case 1: one to one match for isolated sources. Case 2: two stars have the same match which in most cases it is a duplicated source. Case 3: the best match is selected from a neighbourhood. Case 4: wrong match possibly because the right one is missing.}
\label{xmplot}
\end{figure}

\begin{figure*}
\includegraphics[width=23cm, height=6cm, keepaspectratio]{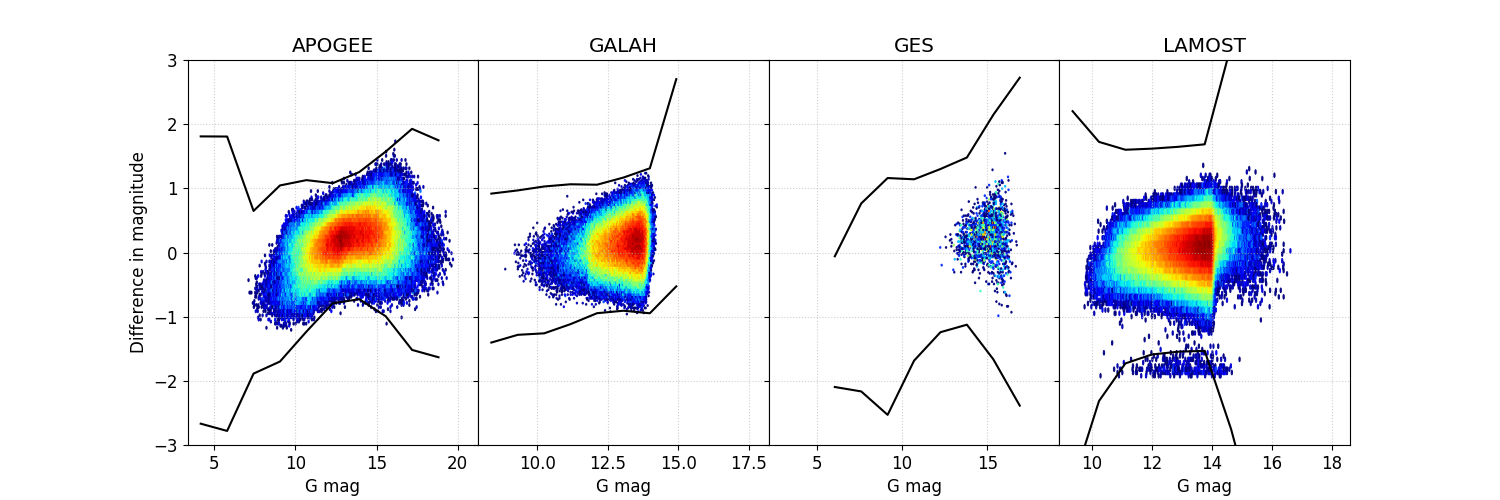}
\caption{G magnitude difference between {\em Gaia} match and the surveys: APOGEE, GALAH, GES, LAMOST, and RAVE respectively. The plots are color coded to the stellar number density. The horizontal black lines indicate the $\pm$3\,$\sigma$ threshold for outliers per magnitude bin (10 bins in total).}
\label{noneighbourhood}
\end{figure*}

An essential part of this work is to find which stars from the spectroscopic surveys are the counterpart sources in {\em Gaia}. The algorithms for the cross match (XM) of the {\em Gaia} DR2 astrometric data with external catalogues are described in \cite{Marrese2017, Marrese2019}\footnote{https://gea.esac.esa.int/archive}. The defined algorithms are positional and exploit the enormous number of {\em Gaia} sources with accurate positions, proper motions, and parallax measurements on an object-by-object basis. The XM is not only a source-to-source problem but also a local one, meaning that the neighbourhood around the possible match is investigated by assigning probabilities to all neighbours based on their angular distances, but also on the local surface density of the external catalogue to choose the best neighbour among them.

The XM algorithm is slightly different for large \textit{dense} surveys, and for \textit{sparse} catalogues. The ground-based spectroscopic surveys in this work are sparse, meaning that the XM treats them as the leading catalogues and {\em Gaia} as the second. In this case, a given object in the leading catalogue is matched with all nearby objects in the second catalogue whose position is compatible within position errors with these nearby targets defined as \textit{neighbours}. When a single neighbour is found, it is the counterpart, i.e. best neighbour (case 1 in Fig.~\ref{xmplot}). When more than one neighbour is found, the best neighbour is selected according to a figure of merit (case 3 in Fig.~\ref{xmplot}). 

If two or more objects from the leading catalogue are matched to the same object in the second, then these sources are referred to as \textit{mates}. For sparse catalogues as the spectroscopic ones in this work, mates are not allowed because a one-to-many match is forced since {\em Gaia} has a higher spatial resolution. Blends therefore, are not expected as opposed to dense catalogues where {\em Gaia} with its higher resolution is the leading catalogue. Mates for sparse catalogues are usually duplicate sources from repeated measurements and are labelled with the same identification in the leading catalogues (case 2 in Fig.~\ref{xmplot}). APOGEE, LAMOST, and RAVE have reported duplicates but GALAH and GES provide only unique sources. There are cases however, in which two or more objects have the same {\em Gaia} match but are not identified as the same source in the original catalogue. This implies that either the survey mislabelled one of the sources and they actually are duplicates (case 2 in Fig.~\ref{xmplot}), or there is a missing source in {\em Gaia} and one of the objects has no counterpart (case 4 in Fig.~\ref{xmplot}). We expect the majority of the mates found nevertheless to be mislabelled duplicates.

We assess the cases where the XM has found mates and investigate if these sources are in fact duplicates or problematic matches (case 2 in Fig.~\ref{xmplot}). While the XM considers five parameters (right ascension, declination, parallax, and proper motions), we utilise here additional information on the magnitudes, and radial velocities for the classification of the possible duplicates. For each pair of possible duplicates, we calculate their differences in magnitudes and in radial velocities to determine the Mahalanobis distance \citep[MD,][]{Mahalanobis1936} of the above parameters as a metric for defining outliers. The MD is a multiple regression generalization of one-dimensional Euclidean distance and is described by:
\begin{equation}
    MD^2 (X_i)=(X_i-\bar{X})^T\hat{V}^{-1}(X_i-\bar{X})
\end{equation}
where X$_{i}$ are the paired differences in magnitude, and paired differences in radial velocities, $\bar{X}$ is the mean values of the above parameters, and $\hat{V}$ is the covariance matrix. If the MD for a pair is smaller than a threshold, then the mates are in fact the same star, i.e. a duplicate. On the other hand, any outliers found indicate a possible mismatch and are flagged. The distribution of the MD$^{2}$ is known to be chi-squared ($\chi^{2}_{n}$) with n degrees of freedom, n=2 in this case \citep{Gnanadesikan1972}. Then, the adopted rule for identifying outliers is the \% quantile (q) of the $\chi^{2}_{n=2}$ and observations are flagged as outliers if MD$^{2}$ $>$ $\chi^{2}_{n=2}(q)$.

We confirm that most of mates identified from the XM are indeed duplicates (see Table~\ref{table:1}) and we update their identification name in each survey to be treated as duplicates for the rest of the analysis. The initial identification is also kept in a different column for completeness. As a threshold we used different criteria for the cut-off: the 99\% quantile, the 97.5\% quantile, and a 3\,$\sigma$ (standard deviation) as a more conservative cut-off. Even for the most conservative case, 82-97\% of the mates appear as duplicates in all surveys. The problematic matches are flagged as \texttt{mismatches} in a separate column (\texttt{flag\_xm}) and also stars we could not provide a result due to lack of any of the parameters (either magnitude or radial velocity) for the calculation of the MD are flagged as \texttt{possible\_mismatch}. For this analysis, we selected the 97.5\% cut-off which appears in the middle of the three. The numbers of true duplicates and problematic matches are shown in Table~\ref{table:1}. 

\subsection{The XM evaluation}

As we mentioned before, during the XM, the spectroscopic catalogues are used as the leading ones and {\em Gaia} as the second where a one-to-many match is forced. Because this process evaluates the environment, the XM defines for a given object in the leading catalogue, neighbours as nearby objects with positions compatible within errors in {\em Gaia} and are provided in a separate table for checking purposes. When more than one neighbour is found, the best neighbour, i.e. the most probable counterpart according to a figure of merit, is chosen among them. Even though in most cases this process is flawless, there are some reasons why the selection of the best match can be problematic. For instance, in cluster regions the counterpart may not exist in {\em Gaia} and another star close enough could be matched instead (case 4 in Fig.~\ref{xmplot}).

We therefore, want to further evaluate the selection of the best match derived from the XM algorithm by using additional parameters than positional. We divide each catalogue into two categories: \textit{i}) stars with only one match to {\em Gaia}, i.e. star with no neighbours (case 1 in Fig.~\ref{xmplot}) or \textit{ii}) stars where the {\em Gaia} match is selected from a neighbourhood of possible matches (case 3 in Fig.~\ref{xmplot}).

\subsubsection{Stars with no neighbourhood}

First, we analyse stars which are assigned to a unique match in {\em Gaia}. This category comprises of the majority of the stars in each catalogue (95\% for APOGEE, 98\% for GALAH, 93\% for GES, and 97\% for LAMOST). We use photometry, namely the magnitudes, to verify the XM selection. We convert the magnitudes of each survey to the G mag system using the conversion functions of \cite{Evans2018} for the range of magnitude applicability.

If the match is correct, then the G mag from {\em Gaia} DR2 source should agree with the converted G mag from the source in the leading catalogue within 3\,$\sigma$. We show the difference in G magnitudes between the {\em Gaia} best match and the survey object in Fig.~\ref{noneighbourhood}. We select the $\sigma$ to be magnitude dependent (calculated per bin) so that we will not exclude the wings of the distribution. We assume that the stars outside the 3\,$\sigma$ (black lines) in Fig.~\ref{noneighbourhood} are the outliers due to the mismatch. The magnitude comparisons reach up to significant differences, typically higher than the expected ones from the photometric catalogues where the magnitudes are taken. Apart from the mismatches from the cross match of the catalogues, other effects could play a role here so their identification should be taken with caution. In particular, here we compare the G mag converted from photometric transformation of the spectroscopic catalogues with a precision of the transformation set by the $\sigma$ of 0.369\,mag \citep{Evans2018}. We thus, expect some stars such as variable stars or stars with high photometric errors naturally to show high differences.  

Nevertheless, the fraction of the possible mismatches from the stars with available photometry is very small:  0.9\% for APOGEE, 0.5\% for GALAH, 0.8\% for GES, 1.4\% for LAMOST, and 0.6\% for RAVE. For these stars, we raise a flag and assign them as mismatches ($\texttt{flag\_xm}$ = $\texttt{mismatch}$). 

\subsubsection{Stars with neighbourhood}

The number of stars with multiple neighbours in {\em Gaia} is less than 7\%, occurring mainly in crowded fields such as in the bulge or in stellar clusters. We investigate if the best match selected among the neighbours is indeed the correct one, or if any of the neighbours in {\em Gaia} could be better suited. We calculate the G magnitudes of the sources in the leading catalogues using the same conversions as previously. 
If the difference in G mag between the survey star and the neighbour in {\em Gaia} is smaller than of the best match, then the neighbour is now considered the best. In these cases, we flag the stars as mismatches ($\texttt{flag\_xm}$ = $\texttt{mismatch}$). The fraction of the possible mismatches from the stars with neighbourhood is comparable to the previous test: 4.1\% for APOGEE, 4.9\% for GALAH, 4.6\% for GES, and 4.9\% for LAMOST.

The results of the above tests show a robust confirmation of the efficiency of the XM. We are confident that most of our stars are matched properly with the {\em Gaia} DR2. 

\section{The RV error analysis}\label{errors}

In this Section, we normalize the internal RV errors of each survey to make them homogeneous across surveys. Homogenizing the errors allows us to combine the surveys using weighted averages to derive final RVs for stars observed in more than one survey. This process is necessary because each survey has heteroscedastic errors which should infer its precision and accuracy. For surveys with duplicated sources available (APOGEE, RAVE, LAMOST), we use the repeated RV measurements to evaluate the internal errors, while for surveys with few or no duplicates ({\em Gaia}, GALAH, GES), we use the three-cornered hat method. Homogenization of the errors is not an easy task because we have to account for the systematic and random errors of each survey. We focus here on the random component of the errors since they can be treated statistically as opposed to the systematic. 

\subsection{Error normalization from repeated measurements}

\begin{table*}
\caption{Statistics for the paired RV differences from the repeated measurements: Mean of the RV differences, $\sigma$ is the standard deviation, normalized $\sigma$ is derived from the $\sigma$ divided to the square root of the sum of quadratic errors, Median of the RV differences, MAD is the median absolute deviation, normalized MAD is derived from the MAD divided to the square root of the sum of quadratic errors, and N is the number of paired differences. }
\centering
\begin{tabular}{l c c c c c c r}
\hline\hline
\multirow{3}{*}{Survey} & \multirow{2}{*}{Mean} & \multirow{2}{*}{$\sigma$} &  normalized & \multirow{2}{*}{Median} & \multirow{2}{*}{MAD} & normalized & \multirow{3}{*}{N} \\
                        &                       &                           &   $\sigma$  &                         &     & MAD                         &   \\
       & (km\,s$^{-1}$) & (km\,s$^{-1}$) & (km\,s$^{-1}$) & (km\,s$^{-1}$) & (km\,s$^{-1}$) &  (km\,s$^{-1}$) &  \\
\hline
   APOGEE  & --0.02 & 5.26 & 6.8 & 0.00   & 0.23 & 5.9 & 43\,596  \\
   GES     & --0.13 & 3.29 & 1.2 & --0.16 & 0.63 & 1.0 & 31     \\
   RAVE    & 0.09   & 7.36 & 1.1 & 0.09   & 1.42 & 0.8 & 102\,545  \\
   LAMOST  & --0.20 & 8.74 & 0.5 & --0.09 & 3.42 & 0.4 & 1\,893\,154 \\
\hline
\end{tabular}
\label{table:2}
\end{table*}

\begin{figure}
\includegraphics[width=9.0cm, clip]{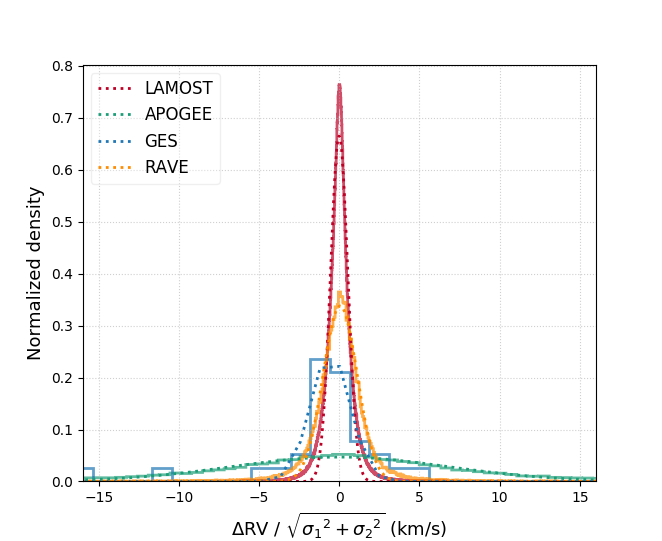}
\caption{Distributions of the RV paired differences of the stars with multiple measurements in each survey normalized to their errors (solid lines). The Gaussian fits are plotted with dotted lines. The statistics of these distributions are given in Table~\ref{table:2}.}
\label{duplicates}
\end{figure}

\begin{table}
\caption{Summary of the error normalization factor for all surveys. The last column indicates the method used: dupl is from repeated measurements, TCH is the three cornered-hat method, and avg is the average of both. }
\centering
\begin{tabular}{l c c}
\hline\hline
Survey & factor & Method \\
\hline
   APOGEE & 5.9 & dupl \\
   {\em Gaia}   & 1.5 & TCH  \\
   GALAH  & 2.0 & TCH  \\
   GES    & 0.8 & avg  \\
   RAVE   & 0.8 & dupl \\
   LAMOST & 0.4 & dupl \\
\hline
\end{tabular}
\label{errorfactor}
\end{table}

The analysis of the repeated RV measurements will give us insights on the precision of each survey. We have a satisfactory number of duplicates from APOGEE, RAVE, and LAMOST for which we calculate the paired differences of the repeated measurements and show the corresponding statistics in Table~\ref{table:2}. GES has only 62 multiple observations and GALAH has none. These surveys will be treated differently in the next Section but we also include GES here for comparison. The results in Table~\ref{table:2} arise from basic filtering we mention in Sect.~\ref{datasources} which affects mainly APOGEE and RAVE. APOGEE and GES exhibit the smallest Median Absolute Deviation (MAD) and therefore, the highest precision which is expected as these surveys operate in higher resolution compared to RAVE and LAMOST. In fact, the MAD decreases with increasing spectral resolution. 

If we assume that each observed RV is a random measurement, it will follow a Gaussian distribution centered on the true RV with dispersion given by the RV uncertainty. The difference between two repeated independent measurements, $\Delta$RV = RV$_{1}$ -- RV$_{2}$ with $\sigma_{RV_{1}}$, $\sigma_{RV_{2}}$ their respective errors, follows a Gaussian distribution centered on zero with a dispersion given by $\sigma_{RV}$ = $\sqrt{\sigma_{RV_{1}}^{2} + \sigma_{RV_{2}}^{2}}$. In case the RVs and their related uncertainties are well determined, the distribution of $\Delta$RV normalised to their errors should be a Gaussian with zero mean value and dispersion of one. Any deviation from unity could mean that the errors are over- or under-estimated. Figure~\ref{duplicates} shows the normalized $\Delta$RV distributions for the repeated measurements. We select the normalized MAD from Table~\ref{table:2}, less sensitive to outliers, as a weight unit factor (normalization factor) to multiply to the survey errors for their normalization hereafter. The MAD is the best choice in our case because we do not filter strictly the surveys for outliers and therefore, we expect to have RVs that deviate a lot from the exact RV value with the goal to correct them in the proceeding steps. GES and RAVE have MAD around unity but LAMOST has overestimated errors by a factor of $\sim$2.5. APOGEE, on the other hand, has underestimated errors by a factor of $\sim$6. The underestimation for the APOGEE errors is also demonstrated in \cite{Cottaar2014} but in their case is estimated by at least a factor of 3. We note that the distributions are not fully Gaussian but have extended tails represented better by Cauchy–Lorentzian distributions (see Gaussian fits in Fig.~\ref{duplicates}). 
Even though we have an estimation of the error realisation for GES, it may not be representative for the whole survey as it only relies on 31 $\Delta$RV measurements for 62 stars. 

\subsection{Error normalization from the three-cornered-hat method}

To estimate the random errors for the surveys with few or no duplicates, such as GES, GALAH, and {\em Gaia}, we use the three-cornered-hat (TCH) method. This method has been developed to investigate the frequency stability of atomic clocks \citep[e.g.,][]{Gray1974}, and was applied for noise analyses of various data, in particular, astronomical and geodetic time series \citep[e.g.,][]{Malkin2013}. The TCH method is applied to three independent data sets, in our case to different catalogues, which are described by the following system assuming they are uncorrelated:
\begin{equation}
 \sigma^{2}_{12} = \sigma_{1}^{2} + \sigma_{2}^{2}, \\
 \sigma^{2}_{13} = \sigma_{1}^{2} + \sigma_{3}^{2},  \\
 \sigma^{2}_{23} = \sigma_{2}^{2} + \sigma_{3}^{2}  \\
\end{equation}
with the solution:
\begin{align}\label{sigmaeq}
 \sigma^{2}_{1} = (\sigma_{12}^{2} + \sigma_{13}^{2} - \sigma_{23}^{2})/2  \nonumber \\
 \sigma^{2}_{2} = (\sigma_{12}^{2} + \sigma_{23}^{2} - \sigma_{13}^{2})/2  \\
 \sigma^{2}_{3} = (\sigma_{13}^{2} + \sigma_{23}^{2} - \sigma_{12}^{2})/2  \nonumber
\end{align}
where $\sigma_{i}^{2}$ are the unknown RV variances we want to determine for each, \textit{i}, catalogue: $\sigma_{i}^{2}$ = $\sum$ $\frac{(RV_{i} - RV_{true})^{2}}{N}$ with the true RVs also unknown, and $\sigma_{ij}^{2}$ are the observed RV variances of the paired differences between the \textit{i} and \textit{j} catalogues: $\sigma_{ij}^{2}$ = $\sum$ $\frac{(RV_{i} - RV_{j})^{2}}{N}$. Unfortunately, the method can be problematic because it can produce negative variances if the data under investigation are correlated. For the cases we consider here, we did not face this problem which means that convariances are very small. 


To derive the expected $\sigma_{GALAH}$ with the TCH method, we use the combination of surveys with the most stars in common: APOGEE, LAMOST, and GALAH (210 stars). Before we calculate the variances of the RV paired differences ($\sigma^{2}_{ij}$), we correct for the ZP differences between the catalogues. Following Eqs.~\ref{sigmaeq}, we calculate $\sigma_{GALAH}$\,=\,0.23\,km\,s$^{-1}$. To obtain the error normalization factor, this value has to be compared with the formal RV errors of GALAH, which have a median value for this sample of 0.12\,km\,s$^{-1}$. We therefore, demonstrate that the normalization factor of GALAH with the TCH method is 2.0 which means that the errors in this survey are underestimated by this factor.

Similarly for GES, we use the combination of surveys with the highest number of stars in common: APOGEE, RAVE, GES. Unfortunately, the number of stars in common with available RVs is only 28 and corresponds to a $\sigma_{GES}$\,=\,0.38\,km\,s$^{-1}$ with median error of 0.56\,km\,s$^{-1}$. The error normalization factor from the TCH method is therefore 0.7 which is slightly lower than the value we obtain from the duplicates of the previous Section. It is difficult to estimate which of the two methods is more accurate because both are based on small samples. Therefore, we use the average value of both which is 0.8, still close to unity for GES. 

For {\em Gaia}, we have the advantage to have more combinations to infer the normalization factor with adequate number of stars: (APOGEE, LAMOST, {\em Gaia}), (APOGEE, RAVE, {\em Gaia}), (LAMOST, RAVE, {\em Gaia}). Following the same method for each of the three sets, we find $\sigma_{Gaia}$\,=\,(1.03, 0.77, 0.90)\,km\,s$^{-1}$ with their median errors of (0.87, 0.57, 0.48)\,km\,s$^{-1}$ which leads to an average factor of 1.5 used for {\em Gaia}. We summarize the normalization factors for all surveys in Table~\ref{errorfactor}.

\section{The RV homogenization process}\label{rvcalibration}

\begin{table}
\caption{Statistics of the $\Delta$RV (= RV$_{Gaia}$ -- RV$_{survey}$). For each survey we report the number of stars in common, the mean with its standard deviation ($\sigma$) and the median with its MAD, before and after the correction described in Sect.~\ref{rvcalibration}. In this work, we consider the median as the zero point difference of each survey with {\em Gaia}.}
\centering
\scalebox{0.82}{
\begin{tabular}{l c c c c r}
\hline\hline
\multirow{2}{*}{Survey} & Mean $\Delta$RV & Median $\Delta$RV & $\sigma$       & MAD            &  \multirow{2}{*}{N} \\
                        & (km\,s$^{-1}$)  & (km\,s$^{-1}$)    & (km\,s$^{-1}$) & (km\,s$^{-1}$) & \\
\hline
   APOGEE & 0.09 & 0.04 & 4.36  & 0.61 & 177\,410  \\
   GALAH  & 0.61 & 0.49 & 4.40  & 0.76 & 91\,416   \\
   GES    & 0.36 & 0.22 & 4.76  & 0.91 & 1\,211    \\
   RAVE   & 0.28 & 0.32 & 5.24  & 1.23 & 436\,709  \\
   LAMOST & 5.18 & 4.97 & 6.85  & 2.99 & 1\,047\,324 \\
\hline
\multicolumn{6}{c}{After RV correction}\\
\hline
   APOGEE & --0.01 & --0.01 & 3.89 & 0.58 & 174\,840 \\
   GALAH  & 0.04   & 0.00   & 3.97 & 0.73 & 86\,841  \\
   GES    & 0.10   & 0.01   & 4.72 & 0.90 & 1\,195   \\
   RAVE   & --0.05 & --0.01 & 4.50 & 1.07 & 192\,979 \\
   LAMOST & 0.09   & --0.06 & 6.55 & 2.92 & 1\,008\,052 \\
\hline
\end{tabular}}
\label{table:3}
\end{table}

\begin{figure}
\centering
\includegraphics[width=9.0cm, clip]{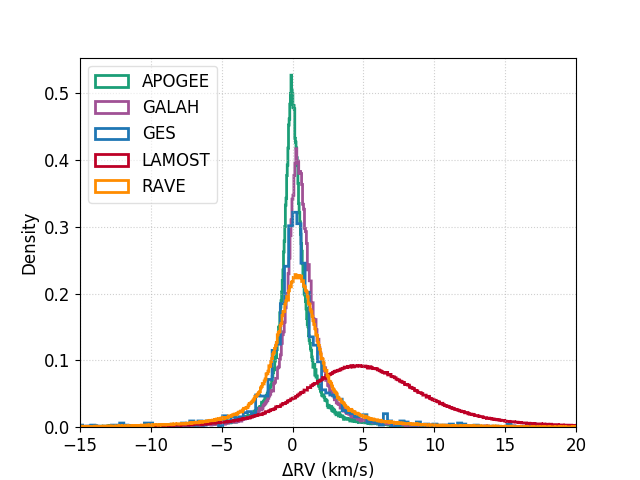}
\caption{Histograms of the RV differences computed as: $\Delta$RV = RV$_{Gaia}$ -- RV$_{survey}$. The statistics of these distributions are given in Table~\ref{table:3}.}
\label{rv_dist}
\end{figure}

\begin{figure*}
\centering
\includegraphics[width=29cm, height=12cm, keepaspectratio]{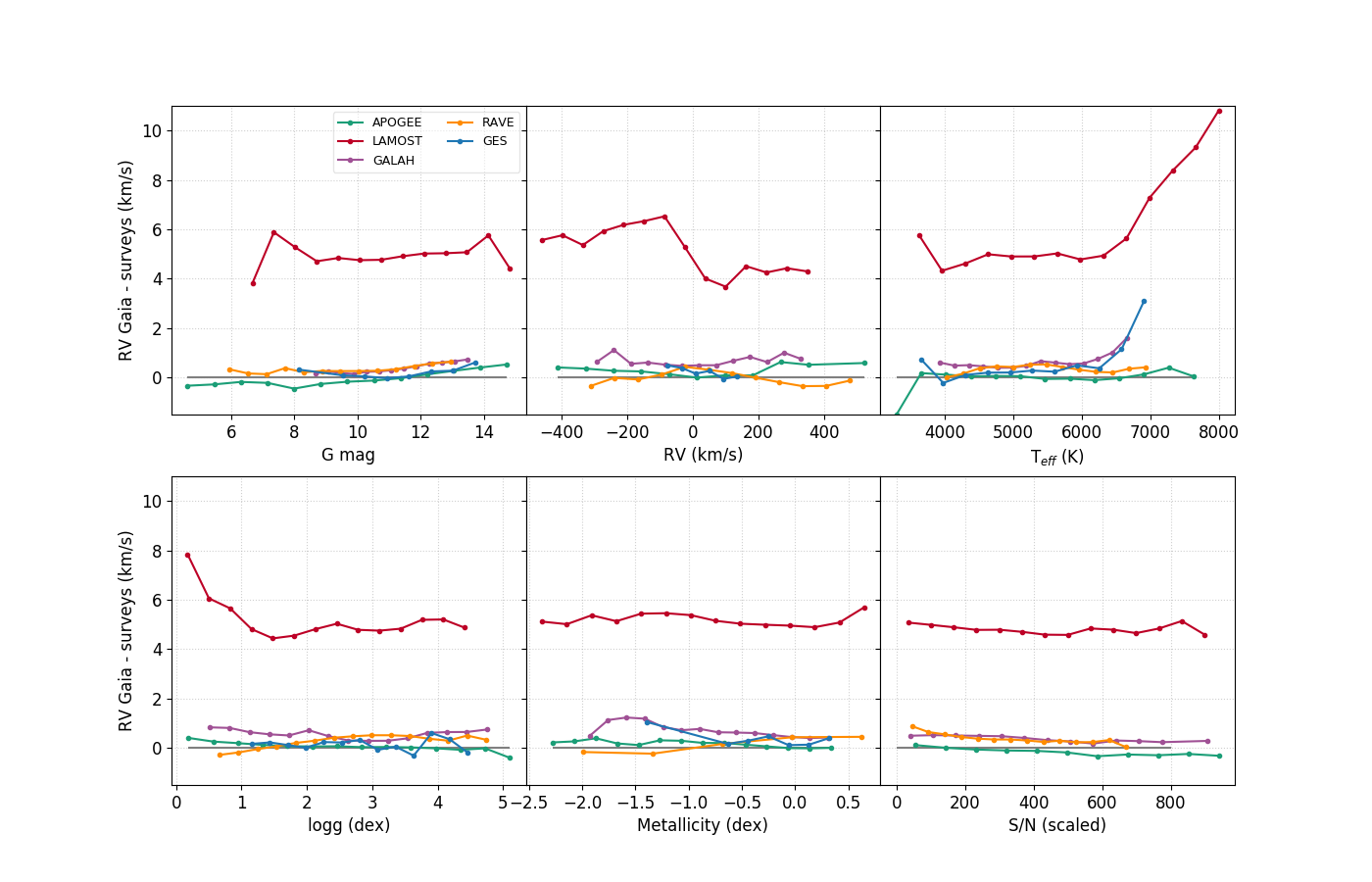}
\caption{The initial RV differences of stars in common with {\em Gaia} as a function of: G magnitude, RV, $T_{\rm eff}$, $\log g$, iron metallicity, and signal-to-noise ratio of the surveys. The S/N is scaled for visual convenience and GES does not provide S/N measurements. The RV differences are binned to contain more than 10 entries for each bin.}
\label{rv_raw}
\end{figure*}

Once the previous steps are concluded, we can compare the RVs of the stars in common with {\em Gaia} and calculate the statistics of their RV differences ($\Delta$RV = RV$_{Gaia}$ -- RV$_{survey}$). The results in Table~\ref{table:3} show the ZP with the highest systematic offset and the highest standard deviation being from LAMOST. We consider the median $\Delta$RV as the ZP in this work. The $\sigma$ and MAD depend on the spectral resolution as expected. The $\Delta$RV distributions are shown in Fig.~\ref{rv_dist}. In this Section, we split our calibration methodology in two parts. First, we calibrate the {\em Gaia} RVs to an arbitrary frame, to show no trends as a function of the main parameters the RVs depend on compared to the higher resolution ground-based surveys (Sect.~\ref{gaia_rv_correction}) and second, we calibrate the ground-based surveys for their intrinsic RV trends (Sect.~\ref{surveys_rv_calibration}).

\subsection{{\em Gaia} RV calibration}\label{gaia_rv_correction}

\begin{table}
\caption{The parameter space covered by all the {\em Gaia} internal RV calibrations computed in Sect.~\ref{gaia_rv_correction}.}
\centering
\scalebox{0.9}{
\begin{tabular}{l c c c c}
\hline\hline
\multirow{2}{*}{Survey} &  G mag & T$_{\rm eff}$ & $\log g$ & [Fe/H] \\
       & (dex)  & (K)           & (dex)    & (dex)  \\
\hline
   APOGEE & 3.3 - 16.0 & 3120 - 8807 & 0.00 - 5.67 & --2.30 - 0.64 \\
   GALAH  & 8.1 - 14.1 & 3832 - 6969 & 0.36 - 4.90 & --2.01 - 0.57 \\
   GES    & 4.3 - 14.8 & 3735 - 7099 & 0.75 - 4.87 & --2.25 - 0.60 \\
   RAVE   & 5.1 - 14.0 & 2998 - 9066 & 0.03 - 5.05 & -             \\
\hline
\end{tabular}}
\label{table:4}
\end{table}

\begin{figure}
\hspace{-0.8cm}
\includegraphics[width=10.5cm, height=20cm, keepaspectratio]{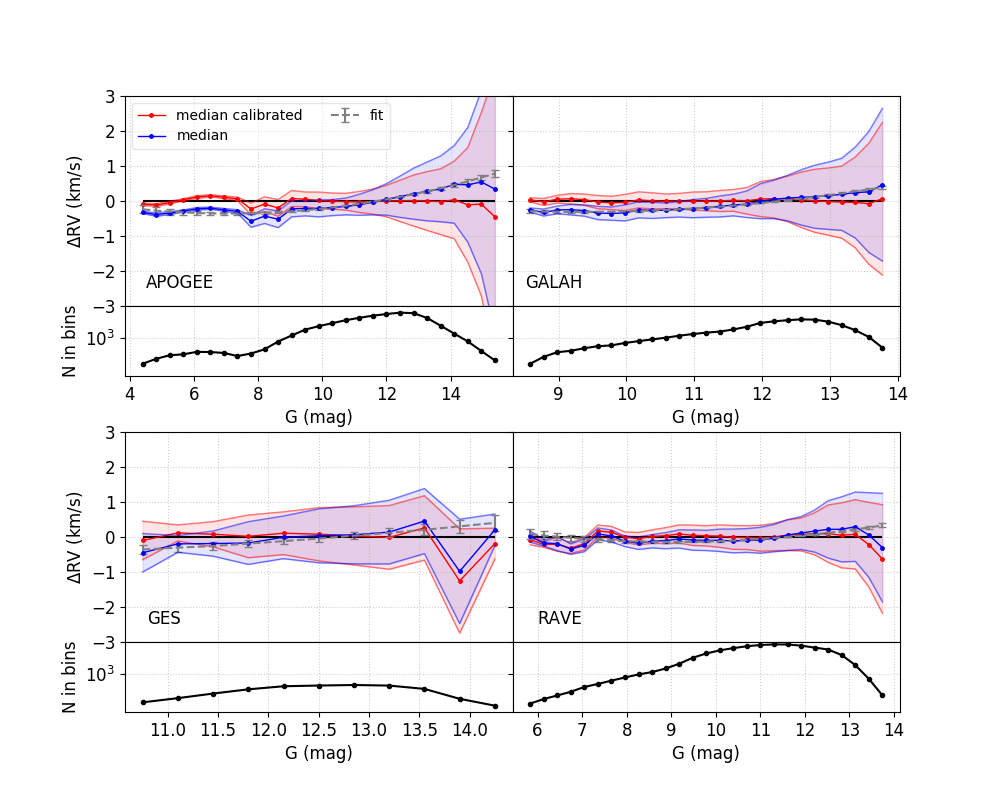} 
\caption{The calibration of {\em Gaia} RVs for stars in common with APOGEE, GALAH, GES, and RAVE as a function of G\,mag derived by fitting a second degree polynomial (Eq.~\ref{rvmaggaiaeq}). The blue and red points represent the median $\Delta$RV of each bin with $>$\,10 entries before and after the calibration respectively. The grey points are the binned $\Delta$RV of the fit. The blue and red shadowed areas are the MAD of each bin. The bottom panels of each plot show the number of stars in each bin.}
\label{rvgaia_mag}
\end{figure}

Figure~\ref{rv_raw} shows the $\Delta$RV as a function of various parameters the RVs depend on. The trends of $\Delta$RV as a function of G mag, metallicity, and $T_{\rm eff}$ are almost identical for most surveys. The $\Delta$RV deviations are higher for fainter stars (G\,$>$\,12 mag) and for hotter stars ($T_{\rm eff}$\,$>$\,6000\,K). We also notice a negative trend with metallicity. In this Section, we investigate these trends assuming that if these correlations are present in most surveys, in particular the ones with the highest resolution, then they must originate from {\em Gaia} and our goal is to eliminate them.

\subsubsection{The G magnitude trend}

Concerning G magnitude, there is a clear trend of $\Delta$RV with G\,mag observed in Fig.~\ref{rv_raw} for all surveys even in the case of LAMOST that shows a large ZP offset and significant deviations at the extremes of the magnitude range. The positive trend of $\Delta$RV as a function of magnitude is also demonstrated by \cite{Katz2019} for the validation of {\em Gaia} RVs in comparison to ground-based surveys, pointing out that the trend begins at G\,$\simeq$\,9\,mag and reaches 0.5\,km\,s$^{-1}$ at G\,$\simeq$\,11.75\,mag. We confirm this and show that the trend extends to fainter stars reaching differences of around 0.7\,km\,s$^{-1}$ (see Fig.~\ref{rvgaia_mag}). We calibrated the trend of the {\em Gaia} RVs with G magnitude, by fitting a second order polynomial with least squares using the normalized errors as weights of the form: 
\begin{equation}\label{rvmaggaiaeq}
    RV_{Gaia} - RV_{survey} = \alpha G mag^2 + \beta G mag + \gamma 
\end{equation}

The selection of this form is empirical and a linear function would not fit properly the faintest stars. For the fitting process, we apply the filtering criteria for the {\em Gaia} DR2 photometry suggested by \cite{Evans2018} and \cite{Arenou2018} based on Eqs.~\ref{gaiafilter}. We fit the $\Delta$RV-G\,mag function with the weighted least squares method for each survey separately and ensure that indeed the trends observed from the different surveys are compatible with each other and thus, these biases can be attributed in the {\em Gaia} data sets (see Fig.~\ref{rvgaia_mag}). Then, we define the global coefficients of Eq.~\ref{rvmaggaiaeq} from their weighted mean to correct all RVs from the entire {\em Gaia} DR2 data set even for the stars that are not in common with the ground-based surveys. We exclude the LAMOST from the process of calculating the global coefficients since this survey is carried out in lower resolution than {\em Gaia} and other trends (see next Section) intrinsic to the survey could weaken the genuine trend in the {\em Gaia} data set. The best-fit and final coefficients of Eq.~\ref{rvmaggaiaeq} are presented in Table~\ref{table:5}. The errors of the best-fit coefficients are the standard errors which correspond to the square-root diagonal values of the covariance matrix of the fit. The correction successfully removes the trend in magnitude when comparing {\em Gaia} with APOGEE, GALAH, GES, and RAVE as shown in Fig.~\ref{rvgaia_mag}. This calibration covers the full {\em Gaia} magnitude range 4-16\,mag (see limits of our calibrations in Table~\ref{table:4}) and does not alter the ZP because the surveys are shifted for their ZP with respect to {\em Gaia} before the fitting process. The errors for the dependent values, $\Delta$RV, are calculated from the covariance matrix including all variances (non-diagonal terms) and covariances re-scaled by the $\chi^{2}$ of the fit \citep[e.g.][]{Bevington1969}. The confidence interval we select is the probability of 95\% to find the true value. Then, the errors for the calibrated {\em Gaia} RVs, $\delta$RV$_{calib}$, arise from quadratic sum of the aforementioned errors and the normalized errors of {\em Gaia} (see Sect.\ref{errors}): $\delta$RV$_{calib}$ = ($\delta\Delta$RV$^{2}$ + $\delta$RV$_{Gaia}^{2}$)$^{1/2}$. We use the same method for the error calculation of the least squares fits throughout this paper. 

\begin{figure*}
\hspace{-1.5cm}
\includegraphics[width=21cm, keepaspectratio]{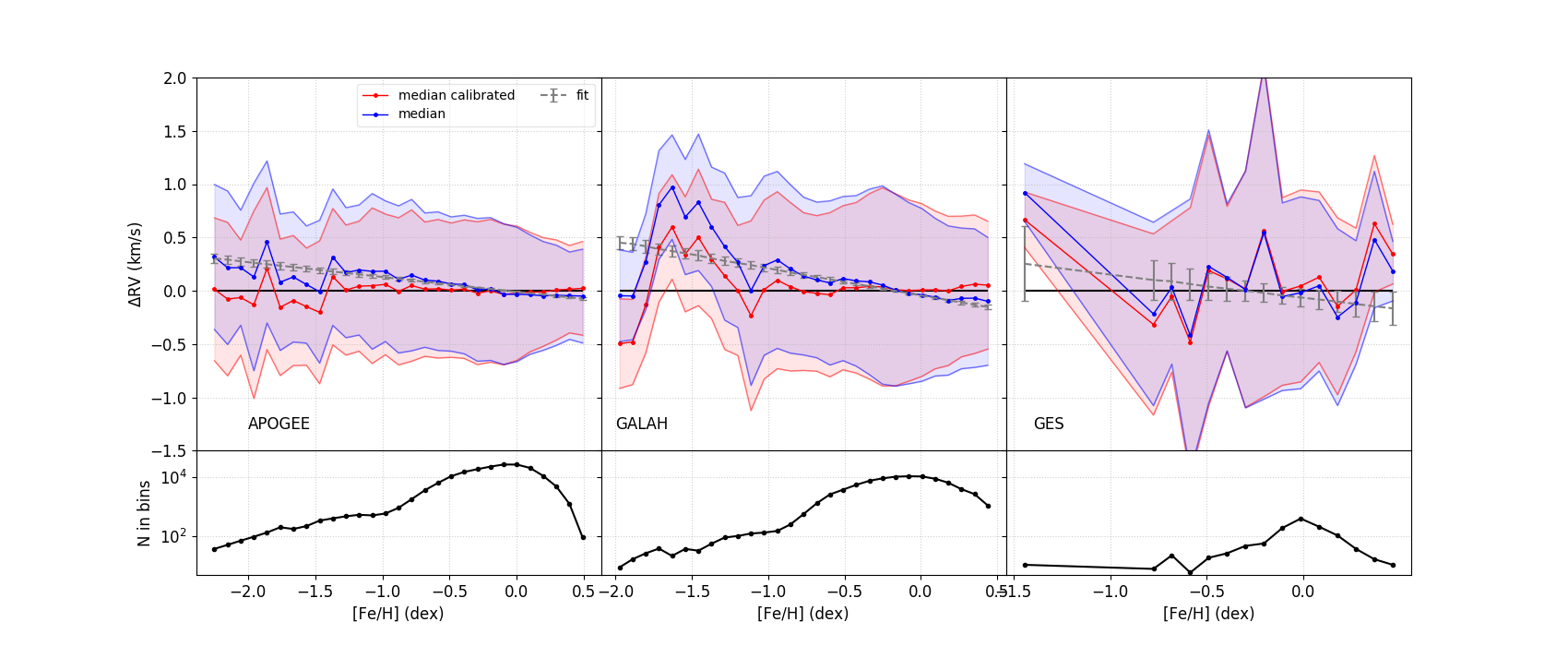} 
\caption{The calibration of {\em Gaia} RVs for stars in common with APOGEE, GALAH, and GES as a function of iron metallicity. The colors and symbols are described in Fig.~\ref{rvgaia_mag}.}
\label{rvgaiacor_metal}
\end{figure*}

\subsubsection{The iron metallicity trend}

Apart from the magnitude trend in Fig.~\ref{rv_raw}, we demonstrate that all surveys apart from RAVE, show a linear correlation of $\Delta$RV with metallicity predominantly from around --2.0 to 0.5\,dex. The $\Delta$RV-$[Fe/H]$ trend is fitted with weighted least squares and has the following form:  
\begin{equation}\label{rvgaiametaleq}
    RV_{Gaia} - RV_{survey} = \alpha [Fe/H] + \beta  
\end{equation}

Interestingly, RAVE shows the opposite trend from the rest of surveys suggesting that this bias is related to RAVE itself. Before applying a linear fit, we have to ensure that metallicities of all surveys are on the same scale because we do not want to propagate errors from the metallicity discrepancies into the {\em Gaia} RVs. We thus compare the metallicities of the surveys in pairs and discover that metallicities for stars in common agree very well within 1\,$\sigma$ ($<$\,0.11\,dex) apart from RAVE which shows a trend\footnote{ We note here that, although good parameters and abundances were later published for RAVE, the initial goal was to derive accurate RVs, not an accurate chemistry for individual stars.}, with $\sigma$ $\simeq$\,0.2\,dex (see Fig.~\ref{metal_rave}). The comparison metallicity plots between the surveys are in Appendix~\ref{metal}. We note that for RAVE we used the iron metallicities from the GAUGUIN pipeline \citep[see references in][]{Guiglion2016} filtered for the clean sample as suggested by the latest release of RAVE \citep{Steinmetz2020b}. We apply a calibration to the RAVE metallicities to scale them with the high resolution surveys (see Appendix~\ref{metal}) but the negative $\Delta$RV-$[Fe/H]$ trend still remains strong, indicating that this trend arises from the RAVE RVs which we will further investigate in the next Section. \cite{Steinmetz2020} also notice this tendency for the RV shift between RAVE and {\em Gaia} DR2 with overall metallicity ($[M/H]$) amounting to 0.5\,km\,s$^{-1}$ differences between metal poor ([M/H]\,$<$\,--1.0\,dex) and metal rich stars ([M/H]\,$>$\,0.0\,dex) which is similar to what we observe here. 

RAVE is not included in the fits to obtain the $\Delta$RV-$[Fe/H]$ calibration coefficients for Eq.~\ref{rvgaiametaleq}. LAMOST is also excluded because it operates in low resolution. The best-fit coefficients of Eq.~\ref{rvgaiametaleq} are presented in Table~\ref{table:5} and the limits of the calibration in Table~\ref{table:4}. For the final coefficients, we use again the mean weighted coefficients shown in Table~\ref{table:5} and present the results in Fig.~\ref{rvgaiacor_metal} for the higher resolution surveys. Even though RAVE and LAMOST are not included for the calculation of the correlation coefficients, we still use their metallicities to calibrate the {\em Gaia} RVs of stars in common with {\em Gaia} from Eq.~\ref{rvgaiametaleq} since we have demonstrated that their metallicities are in agreement with the other surveys as well. 

The metallicity trend is also observed in the {\em Gaia} DR2 validation of the radial velocities for RAVE but not for APOGEE \citep{Katz2019} as we see in this work. A reason may be that in their comparison there is a lack of metal poor stars in common with APOGEE (their figure~12).  

\begin{figure}[t]
\centering
\includegraphics[width=9.5cm, clip]{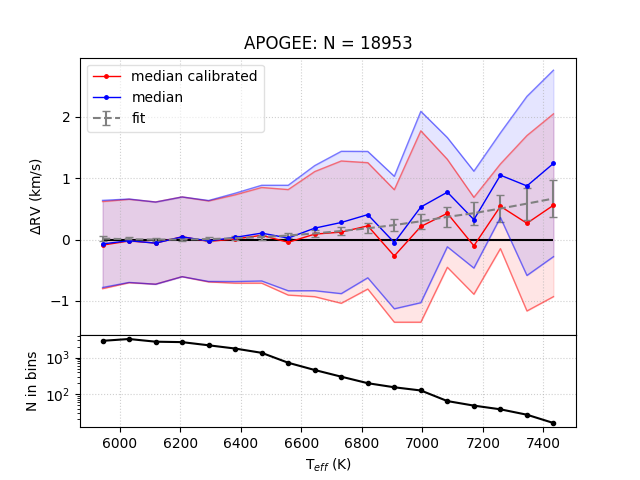}
\caption{The calibration of {\em Gaia} RVs as a function of T$_{\rm eff}$ for APOGEE stars in common with {\em Gaia} to calibrate the {\em Gaia} RVs. The colors and symbols are described in Fig.~\ref{rvgaia_mag}.}
\label{rvgaiacor_teff}
\end{figure}

\subsubsection{The $T_{\rm eff}$ trend}

The last trend we notice is the $\Delta$RV increase as a function of effective temperature for $T_{\rm eff}>$\,5900\,K for all surveys but with different rates (see upper right panel of Fig.~\ref{rv_raw}). This correlation is more difficult to investigate because the different $T_{\rm eff}$ scales of the surveys could play a role. Moreover, since hotter stars have fewer lines in their spectra, their parameters are more difficult to be precisely obtained via spectroscopy than solar-type stars for instance. These stars could also exhibit higher rotational velocities impacting the determination of both their RVs and $T_{\rm eff}$. 

To investigate to what extend this trend arises from {\em Gaia} RVs, it is important to check first if the $T_{\rm eff}$ scale among the surveys is compatible and if the $\Delta$RV-$T_{\rm eff}$ appears when comparing the ground-based surveys themselves. From a comparison of the $T_{\rm eff}$ between the surveys, we notice that their differences are significant for the lowest resolution surveys, RAVE and LAMOST (see Appendix~\ref{metal}). The $T_{\rm eff}$ scale between APOGEE and GALAH is in agreement. Moreover, by comparing the RVs of APOGEE and GALAH as a function of $T_{\rm eff}$, they appear flat. For GES, we do not have enough stars in common for the hotter stars to draw conclusions. The fact that APOGEE and GALAH agree on the temperature scale and their $\Delta$RV-$T_{\rm eff}$ plots are flat but in comparison to {\em Gaia} they show deviations, it is an indication that there could be actually some dependence between {\em Gaia} RVs and $T_{\rm eff}$. In case of APOGEE, the $\Delta$RV reaches 0.5\,km\,s$^{-1}$ and for GALAH around 1.0\,km\,s$^{-1}$ for the hotter stars.

Because the accuracy of the $T_{\rm eff}$ for the hotter stars is low, we follow a conservative approach here. We apply a second degree polynomial to account for this bias but using only APOGEE stars: 
\begin{equation}\label{rvgaiateffeq}
    RV_{Gaia} - RV_{survey} = \alpha T_{\rm eff}^2 + \beta T_{\rm eff} + \gamma 
\end{equation}
This is equivalent to assuming that the APOGEE scale is the true one, and that the $T_{\rm eff}$ trends are caused by {\em Gaia} can thus be estimated from the comparison with APOGEE. If this assumption is wrong, we would have biased the SoS RV behaviour as a function of $T_{\rm eff}$ by up to 0.5\,km\,s$^{-1}$ for the hot stars (but see Sect.\ref{technical}, where we compare with external catalogues).

The best-fit coefficients of Eq.~\ref{rvgaiateffeq} are shown in Table~\ref{table:5} and their limits in Table~\ref{table:4}. The results of the APOGEE calibration are presented in Fig.~\ref{rvgaiacor_teff}. The coefficients of Table~\ref{table:5} are used to correct the {\em Gaia} RVs using the $T_{\rm eff}$ of each survey. The effective temperature of LAMOST shows significant discrepancies compared to the other surveys with same trends, therefore, we correct for it with a linear polynomial shown in Fig.~\ref{teff_comparison}. The calibrated $T_{\rm eff}$ scale for LAMOST is used in Eq.~\ref{rvgaiateffeq} to calibrate the RVs for stars in common with {\em Gaia}.

We point out that the {\em Gaia} RVs in DR2 are computed for stars in the $T_{\rm eff}$ range of 3550-6900\,K because of degraded performance of the RVs outside this range and the restricted grid of templates. Therefore, the deviations for the hotter stars we see here are expected. \cite{Katz2019} mention as well that the {\em Gaia} RV precision is a function of $T_{\rm eff}$ among other parameters in the same direction we see here. Another reason for such trend can be attributed to the increasing dominance of broad Paschen lines in emission in the Calcium triplet region of {\em Gaia} and RAVE. 

The final errors of the {\em Gaia} RVs come from the propagation of the three individual corrections of the fits quadratic summed to the original error. 
 
\subsection{Surveys RV calibration}\label{surveys_rv_calibration}

\begin{figure*}
\centering
\includegraphics[width=29cm, height=12cm, keepaspectratio]{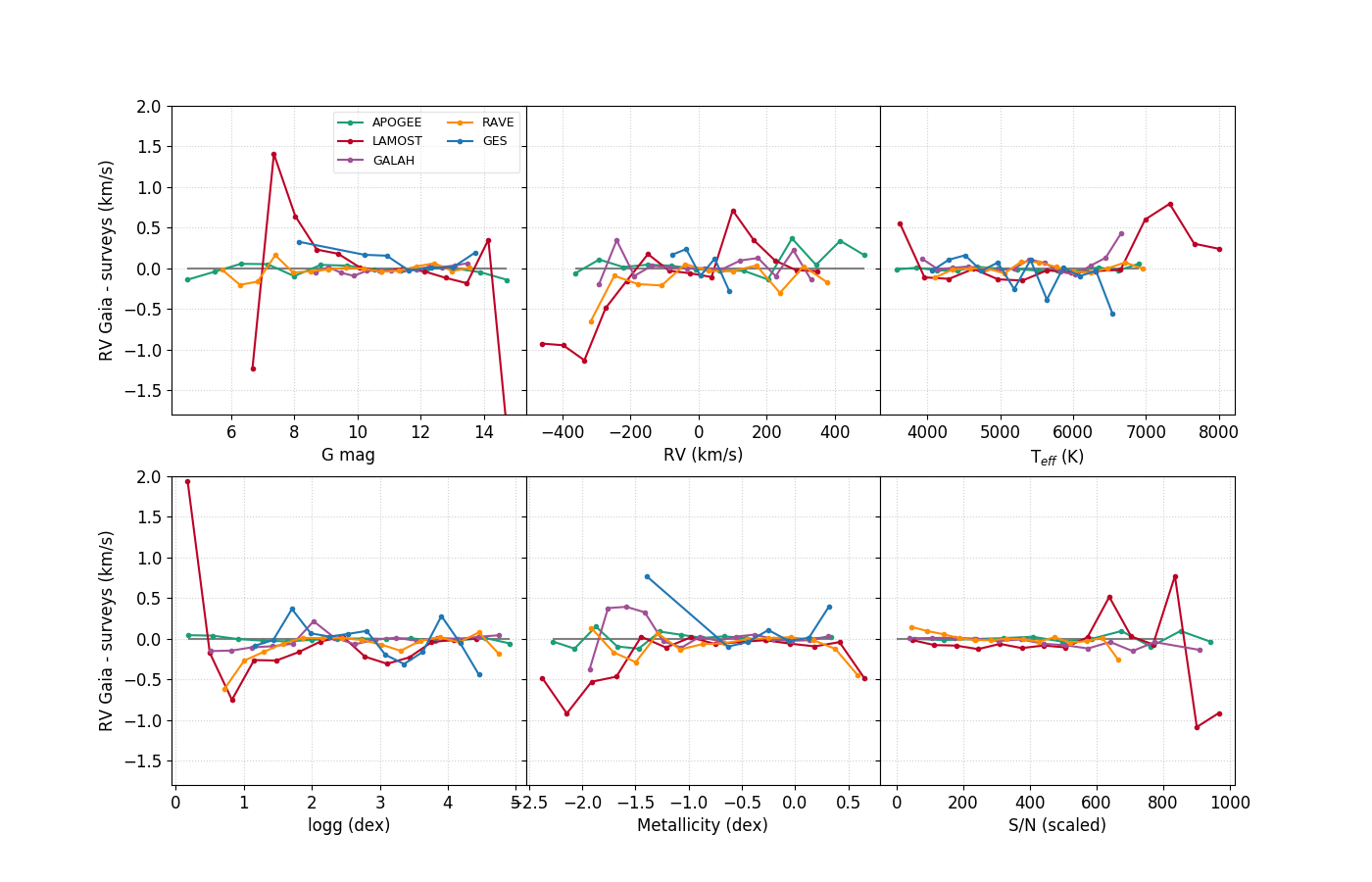}
\caption{The calibrated RV differences of stars in common with {\em Gaia} as a function of: G magnitude, RV, $T_{\rm eff}$, $\log g$, iron metallicity, and signal-to-noise ratio of the surveys. The symbol are the same as Fig.~\ref{rv_raw}.}
\label{rv_calib}
\end{figure*}

\begin{figure}
\centering
\includegraphics[width=9.0cm, clip]{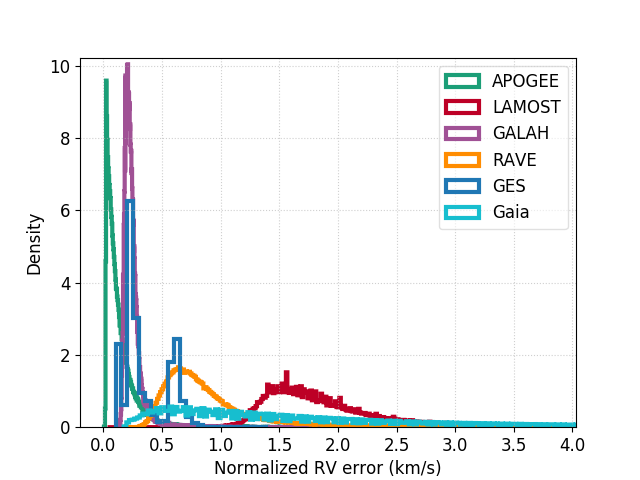}
\caption{The distribution of the normalized errors for all surveys.}
\label{errv_cor}
\end{figure}

After the investigation and correction of the possible biases the {\em Gaia} RVs may suffer, we study the remaining trends of the RV differences between {\em Gaia} and the surveys caused by the internal biases of each survey itself. These trends are more pronounced  for the surveys with lower spectral resolution than {\em Gaia}, such as RAVE and LAMOST. Once the previous dependencies caused primarily by {\em Gaia} are minimized, we apply a multivariate correction of RVs as a function the most relevant parameters: $T_{\rm eff}$, $\log g$, $[Fe/H]$, and S/N to bring these surveys to the corrected {\em Gaia} RV scale. The shape of the function depends on the behaviour of each survey with respect to {\em Gaia}. For instance, GALAH and LAMOST show an evident second order dependence on $T_{\rm eff}$ while the others do not. In the case of LAMOST, it is necessary to add a linear term for RV above and below 100\,km\,s$^{-1}$ (see the top center panel of Fig.\ref{rv_raw}), while the RVs outside these limits are only corrected for the ZP. Finally, GES does not provide unique S/N measurements because their results are based on the star-by-star combination of varying sets of spectra, taken with different instruments (UVES and GIRAFFE) and in different combinations of spectral ranges. We have the following general and empirical function for the RV calibration for all surveys: 
\begin{equation}\label{rvcoreq}
 \Delta RV = \alpha T_{\rm eff}^{2} + \beta T_{\rm eff} + \gamma \log g + \delta [Fe/H] + \epsilon S/N + \zeta RV + \eta
\end{equation}

For a more detailed analysis, we split the stars of the surveys into sub-giants/dwarfs ($\log g$\,$>$\,3.5\,dex) and giants ($\log g$\,$<$\,3.5\,dex). We do that because we notice that dwarfs show a linear correlation in the $\Delta$RV-$\log g$ plot whereas the function for giants appears flatter (see the bottom left panel of Fig.\ref{rv_raw}). The division of the samples into dwarfs and giants at least in the cases of GALAH and APOGEE produced better least-squares fits in terms of their $\chi^{2}$ but there was no significant difference in the overall statistics of their median RV differences. The LAMOST sample was further split to high and low $T_{\rm eff}$ (with the limit of 6200\,K, see the top right panel of Fig.~\ref{rv_raw}) since the temperature correlation could not be fit properly due to the fact that LAMOST contains an enormously larger number of cool stars. The fits are performed for the filtered samples shown in the Appendix~\ref{rvcalibrationappendix} but the correction is applied to the whole surveys with the corresponding coefficients in Table~\ref{table:8}.

The ZP of all surveys is now shifted to the new {\em Gaia} reference. The final corrected statistics for the whole samples are shown in the bottom part of Table~\ref{table:3}. The spreads are now much smaller compared to the initial samples indicating better agreement between the surveys. The comparison of the {\em Gaia} RVs with the calibrated ones from the surveys is shown in Fig.~\ref{rv_calib}. In Appendix~\ref{rvcalibrationappendix}, we also show the paired RV calibrated differences between surveys.

\section{The SoS catalogue}\label{soscatalog}

\begin{table}
\caption{The SoS\,I.}
\centering
\scalebox{0.85}{
\begin{tabular}{l c l}
\hline\hline
Columns              & Units        & Description     \\
\hline
ID                   &              & unique SoS ID\\
RA                   & deg          & Right Ascension\\
Dec                  & deg          & Declination\\ 
RV                   & km\,s$^{-1}$ & RV from the homogenization\\
$\delta$RV           & km\,s$^{-1}$ & error on the RV from the homogenization\\
$\sigma_{RV}$       & km\,s$^{-1}$ & error on the RV considering multiple surveys$^{1}$\\
N                    &              & number of surveys combined to deliver the RV\\
flag\_xm             &              & flag from the XM analysis\\
flag\_rv             &              & flag on the quality of the RV\\
flag\_binary         &              & flag on binarity\\
G                     & mag          & G-band integrated magnitude\\
SurveyID              &              & Surveys from which the SoS RV is derived\\
\hline
\end{tabular}}
\tablefoot{(1) $\sigma_{RV}$ is the weighted error on the RV taking into account the scatter in the measurements when are obtained from different surveys from Eq.~\ref{erroreq}. 
}
\label{table:10a}
\end{table}

\begin{table}
\caption{Auxiliary catalogues to include original data from the catalogues and intermediate products before final homogenization.}
\centering
\scalebox{0.85}{
\begin{tabular}{l c l}
\hline\hline
Columns              & Units        & Description     \\
\hline
SoS ID               &              & SoS ID\\
Survey ID            &              & ID designation for each survey\\
{\em Gaia} ID              &              & {\em Gaia} source ID from DR2 \\
RA                   & deg          & Right Ascension \\
Dec                  & deg          & Declination \\ 
RVcalibrated         & km\,s$^{-1}$ & RV from the homogenization \\
$\delta$RVcalibrated & km\,s$^{-1}$ & error on the RV from the homogenization\\
$\delta RV_{\sigma}$ & km\,s$^{-1}$ & error on the RV from multiple measurements$^{1}$\\
n                    &              & number of multiple measurements\\
flag\_xm             &              & flag from the XM neighbourhood analysis\\
flag\_rv             &              & flag on the quality of the RV\\
flag\_binary         &              & flag on binarity\\
$T_{\rm eff}$        & K            & $T_{\rm eff}$ of the survey\\
$\delta T_{\rm eff}$ & K            & $T_{\rm eff}$ error of the survey\\
$\log g$             & dex          & $\log g$ of the survey\\
$\delta \log g$      & dex          & $\log g$ error of the survey\\
$[Fe/H]$             & dex          & $[Fe/H]$ of the survey \\
$\delta [Fe/H]$      & dex          & $[Fe/H]$ error of the survey\\
\hline
\end{tabular}}
\tablefoot{(1) $\delta RV_{\sigma}$ is the weighted error on the RV taking into account the scatter in the measurements when are obtained from multiple measurements in each survey from Eq.~\ref{erroreq}. }
\label{table:10b}
\end{table}

\begin{figure}
\includegraphics[width=9.5cm, clip]{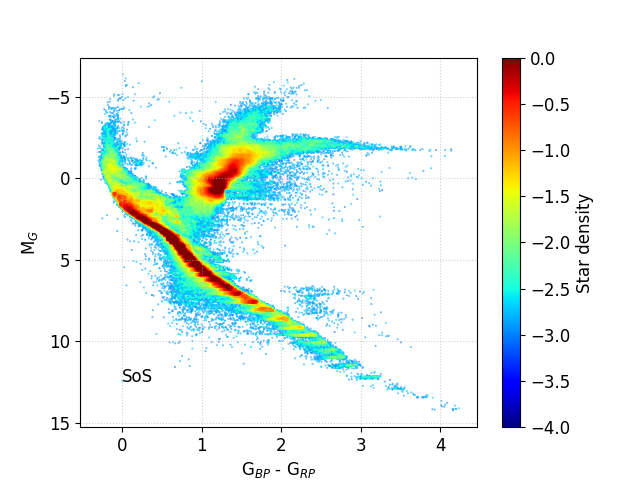} 
\caption{HR diagram for SoS color coded to stellar density in log scale.}
\label{cmd_sos}
\end{figure}

\begin{figure*}
\centering
\includegraphics[width=8.5cm, clip]{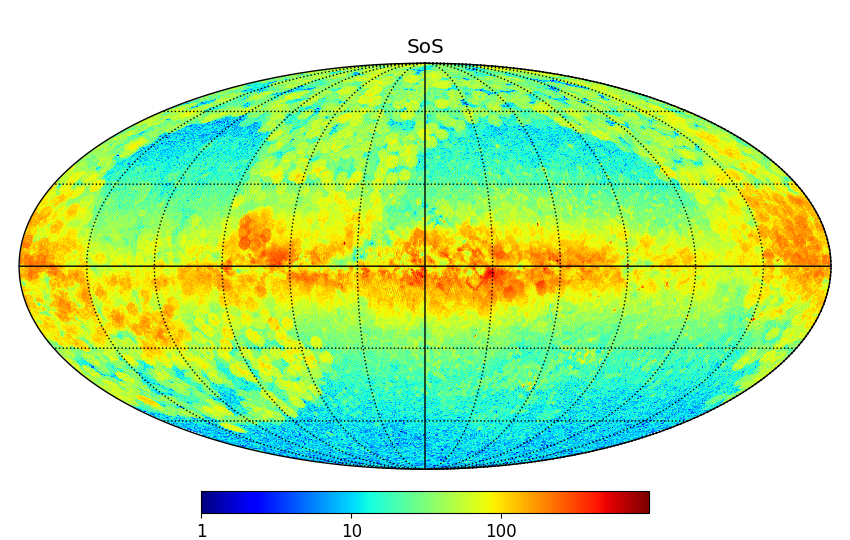} ~~~~~~~~~~~
\includegraphics[width=8.5cm, clip]{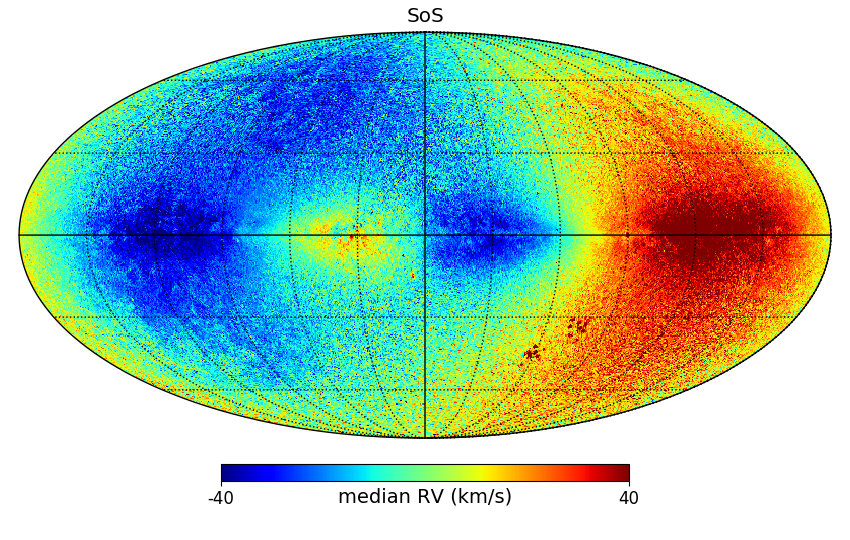}
\caption{Left panel: Surface density distribution in Molleview projection for the stars in SoS in logarithmic scale. Right panel: Same as left panel but color coded to the final SoS RVs (median RV per pixel) for stars with |RV|$<$40\,km\,s$^{-1}$. Both plots are in Galactic coordinates with pixel size of $\sim$\,0.46 degrees.}
\label{sos}
\end{figure*}

\begin{figure}
\hspace{-0.9cm}
\includegraphics[width=10.5cm, clip]{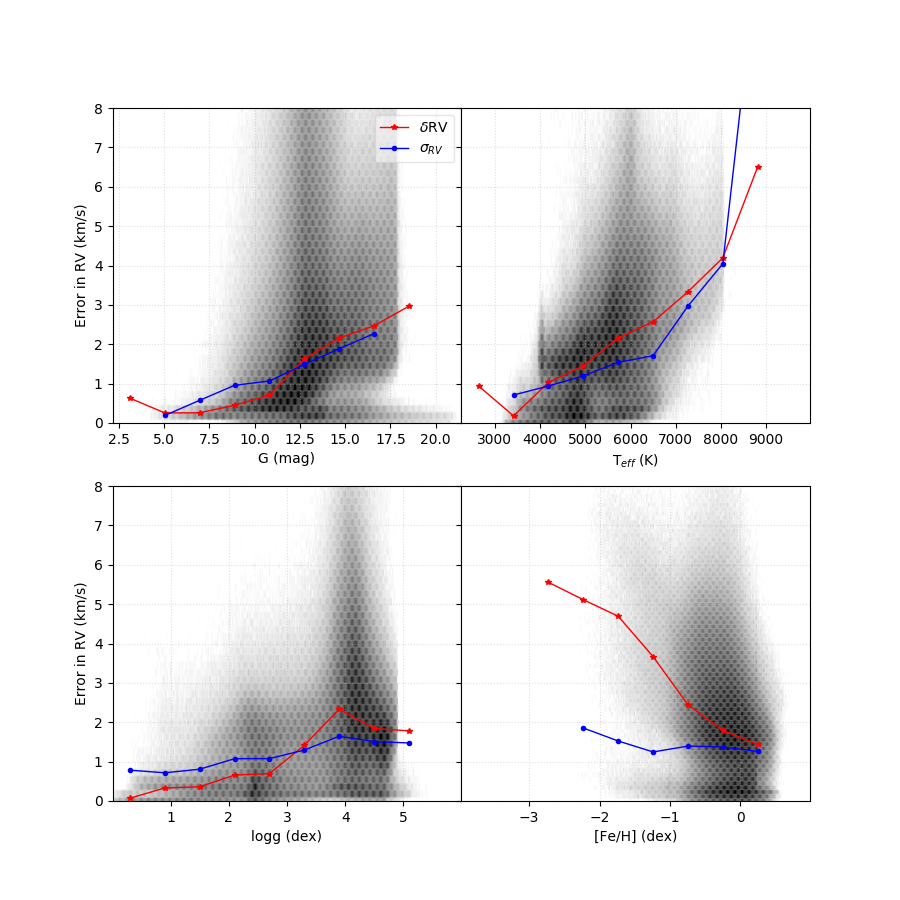} 
\caption{The derived RV errors in SoS after the homogenization process as a function of magnitude and stellar parameters. The errors ($\delta$RV and $\sigma_{RV}$) are binned to their median value. The black background corresponds to the 2D hexagonal binned values of the whole $\delta$RV in SoS. }
\label{errors_sos}
\end{figure}

\begin{figure}
\centering
\includegraphics[width=9.0cm, clip]{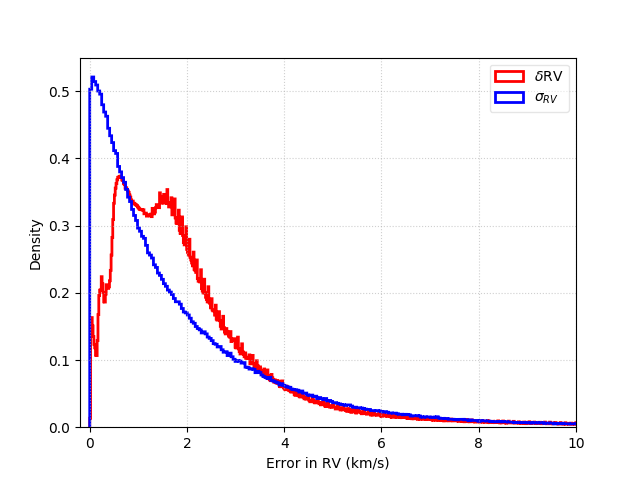} \\
\caption{The distribution of errors ($\delta$RV and $\sigma_{RV}$) in SoS after the homogenization process.}
\label{errors_sos_dist}
\end{figure}

Once we have the calibrated RVs from the previous steps, we merge all surveys together into a unified catalogue in a two step process. First, we get the weighted mean RV and the corresponding weighted error for the duplicate sources in each catalogue to obtain a unique entry for each star in each survey.\footnote{Weighted mean and its error:
\begin{equation}
\overline{x} = \frac{\sum_{i=1}^N w_{i}x_{i}}{\sum_{i=1}^N w_{i}}, w_{i} = \frac{1}{\sigma^{2}_{i}} \nonumber
\end{equation}} 
Second, for stars in common across the surveys, we compute the best RV estimate from a weighted average using the normalized RV errors of each survey as weights. 


The final catalogue contains around 11 million unique stars which amounts to the largest catalogue of RV measurements published so far. In this catalogue, we have combined around 5.1 million stars with RVs from the ground-based spectroscopic surveys and 7.2 million stars with {\em Gaia} RVs. There are only $\sim$52\,000 stars observed solely by the ground-based surveys and do not have a counterpart in {\em Gaia}. Almost half of the SoS (52\%) contains RVs only from {\em Gaia} while the rest is combined with ground-based spectroscopy. 

We provide a quality flag to facilitate the scientific exploration of the SoS catalogue described as \texttt{flag\_rv}:
\begin{itemize}
    \item \texttt{flag\_rv} = 0. These stars have passed the filtering criteria of each survey (Sect.~\ref{datasources}) and have reliable $T_{\rm eff}$, $\log g$, $[Fe/H]$, and available S/N (apart from GES) for all calibrations in the previous Sections. These stars amount to 4.1 million in the SoS ($\simeq$40\%).
    \item \texttt{flag\_rv} = 1. Stars which do not have any of the parameters available for the RV calibration ($T_{\rm eff}$, $\log g$, $[Fe/H]$, and S/N) are shifted only by the ZP to the calibrated {\em Gaia} reference. Also, this flag represents stars with parameters outside the calibration limits. These stars amount to $\simeq$8\% of the SoS. 
    \item \texttt{flag\_rv} = 1.5. The stars with RV measurements obtained only from {\em Gaia} DR2 have their RV calibrated only for their photometric magnitudes and amount to 5.8 million stars. 
    \item \texttt{flag\_rv} = 2. Finally, there are a few stars with problematic RVs as defined from the corresponding surveys ($\simeq$1\%). These stars have lower quality measurements that should only be used with care. We will not use them in the validation and the analysis in the next Sections, but they were included in the catalogue for completeness. Among these, we included 70\,365 sources from {\em Gaia} DR2 that were indicated as potentially contaminated by \cite{Boubert2019}.
\end{itemize}
  
In this first version, SoS contains the basic parameters related to the RV homogenization process as shown in Table~\ref{table:10a}. The calibrated RVs of each survey from the intermediate steps are also kept separately in different tables (see Table~\ref{table:10b}). The distribution of the SoS sources and their RVs are shown in Fig.~\ref{sos}. The star density in SoS is higher for the Galactic center mainly due to {\em Gaia}, for the Galactic anticenter and for the Northern hemisphere mainly due to the LAMOST fields. We also notice overdensities over the Kepler field around (l, b) $\sim$ (76$^{\circ}$, 14$^{\circ}$) which is predominately covered by LAMOST, APOGEE, and {\em Gaia}. The rotation pattern of stars in the Galaxy is evident in the right panel where stars in blue move towards the Sun, whereas stars in red move away in a similar manner presented from the {\em Gaia} DR2 data in \cite{Katz2019}. The centre of the Galaxy is at the centre of the map. The RVs of the for the Small and Large Magellanic Clouds stand out around (l, b) $\sim$ (--57$^{\circ}$, --44$^{\circ}$) and $\sim$ (280$^{\circ}$, --33$^{\circ}$) respectively. In Fig.~\ref{cmd_sos}, we have the final HRD of SoS based on {\em Gaia} DR2 parameters as a function of star density. 

Finally, the errors of the RV homogenization process ($\delta$RV) are shown in Fig.~\ref{errors_sos} as a function of magnitude and stellar parameters indicating the dependence of the RV precision on these parameters. We note that not all SoS stars have available stellar parameters but 99\% of them have G mag measurements. Apart from the weighted RV error, we have also calculated the error on the RV to include the scatter from multiple surveys, $\sigma_{RV}$, used to account for other sources of errors other than systematics \citep[e.g.][]{Malkin2013b}: 
\begin{equation}\label{erroreq}
\sigma^{2}_{RV} = \frac{\sum_{i=1}^N w_{i} (x_{i} - \overline{x})}{(N-1) \sum_{i=1}^N w_{i}}, w_{i} = \frac{1}{\delta RV^{2}_{i}} 
\end{equation} 
where $\overline{x}$ is the average RV for 1 to \textit{i} surveys. These errors are plotted in Fig.~\ref{errors_sos} for stars with RVs derived from two or more surveys. For the brightest stars (G\,$<$12\,mag), the precision is estimated to be better than 1\,km\,s$^{-1}$. The precision worsens sharply for the hot and very metal-poor stars but shows weaker dependence on surface gravity. Dwarf stars show higher errors. The distribution of errors has peaks at 0.05, 0.2, 0.6, and 1.5\,km\,s$^{-1}$ in Fig.~\ref{errors_sos_dist}, depending on the number of multiple measurements and the resolution of the surveys. The $\sigma_{RV}$ is a smoother function peaking at 0.09\,km\,s$^{-1}$ for stars observed in more than one surveys (1.3 million stars), comprising a more robust indicator of the precision in SoS. 

\subsection{Binaries}

\begin{figure}
\hspace{-0.2cm}
\includegraphics[width=10cm, clip]{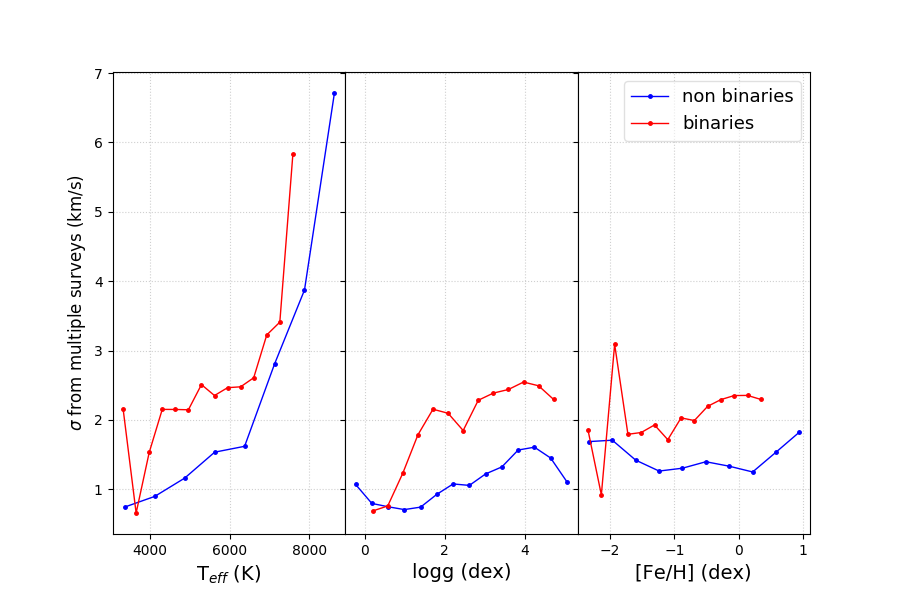} 
\caption{Binaries in SoS. The y axis is the standard deviation in RV from SoS as derived for stars observed from more than one survey. The x axis shows the stellar parameters from the surveys. The red points are the median binned values for the binary population and the blue points the rest of the SoS sample defined here as non binaries.}
\label{sigma_RV}
\end{figure}

Binary systems have an impact on the measured RVs as they are affected by the motion of stars around the center of mass. We have collected information on binarity from the dedicated literature studies for each survey. Whenever a star in the above references is indicated as a binary (or binary candidate), we assigned a separate flag (\texttt{flag\_binary} = 1) otherwise, the flag was set to zero. In particular, \cite{Price-Whelan2020} have detected 19\,635 close-in binaries in APOGEE, \cite{Traven2020} have found 12\,234 double-lined spectroscopic binaries in GALAH, \cite{Merle2017} 641 spectroscopic binaries in GES, \cite{Birko2019} 3\,838 single-lined binary stars in RAVE, \cite{Qian2019} more than 256\,000 spectroscopic binary or variable star candidates in LAMOST, and \cite{Tian2020} $\sim$800\,000 binary candidates for {\em Gaia} DR2. The above stars mount to around 10\% of the SoS and is far from complete as the fraction of binaries in the Galaxy could be between 20\% and 80\%, depending also on spectral type \citep[e.g.][]{Duchene2013}.

We expect these systems to show larger dispersion in their RVs taken at different epochs. Figure~\ref{sigma_RV} shows the standard deviation ($\sigma$) calculated for stars with RVs derived from more than one surveys. The binary population exhibits higher $\sigma$ as a function of the stellar parameters compared to the rest of the sample. Apart from $\sigma$, other parameters in the Gaia data can also indicate binarity, such as the Renormalised Unit Weight Error \citep[RUWE,][see also Appendix~\ref{binaries_appendix}]{Lindegren2018}.

\subsection{Comparison with external catalogues}\label{technical}

\begin{figure}
\hspace{-0.4cm}
\includegraphics[width=10.2cm, keepaspectratio]{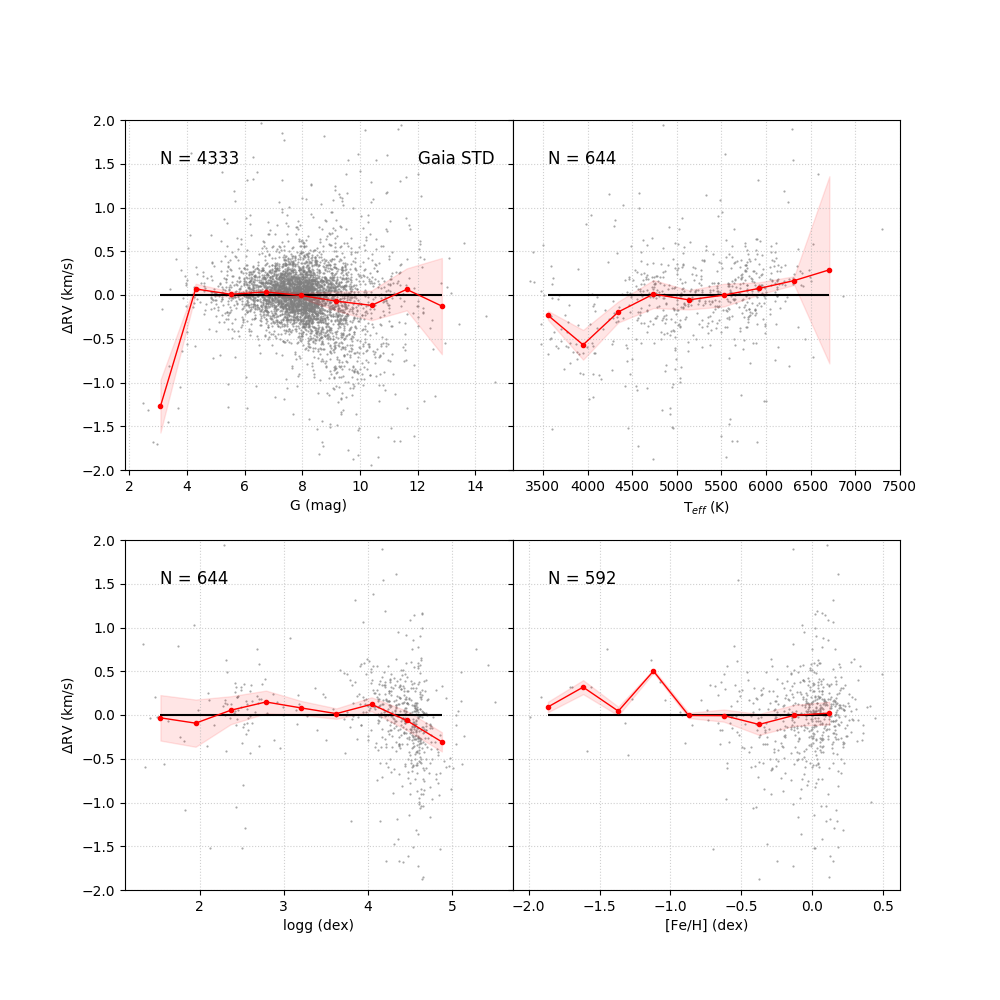} 
\caption{RV comparison between SoS and the {\em Gaia}-STD stars as a function of magnitude, $T_{\rm eff}$, $\log g$, and $[Fe/H]$. The red points are the median binned values and the shadowed area the MAD of each bin. }
\label{gaia_standards}
\end{figure}

\begin{figure}
\hspace{-0.4cm}
\includegraphics[width=10.2cm, keepaspectratio]{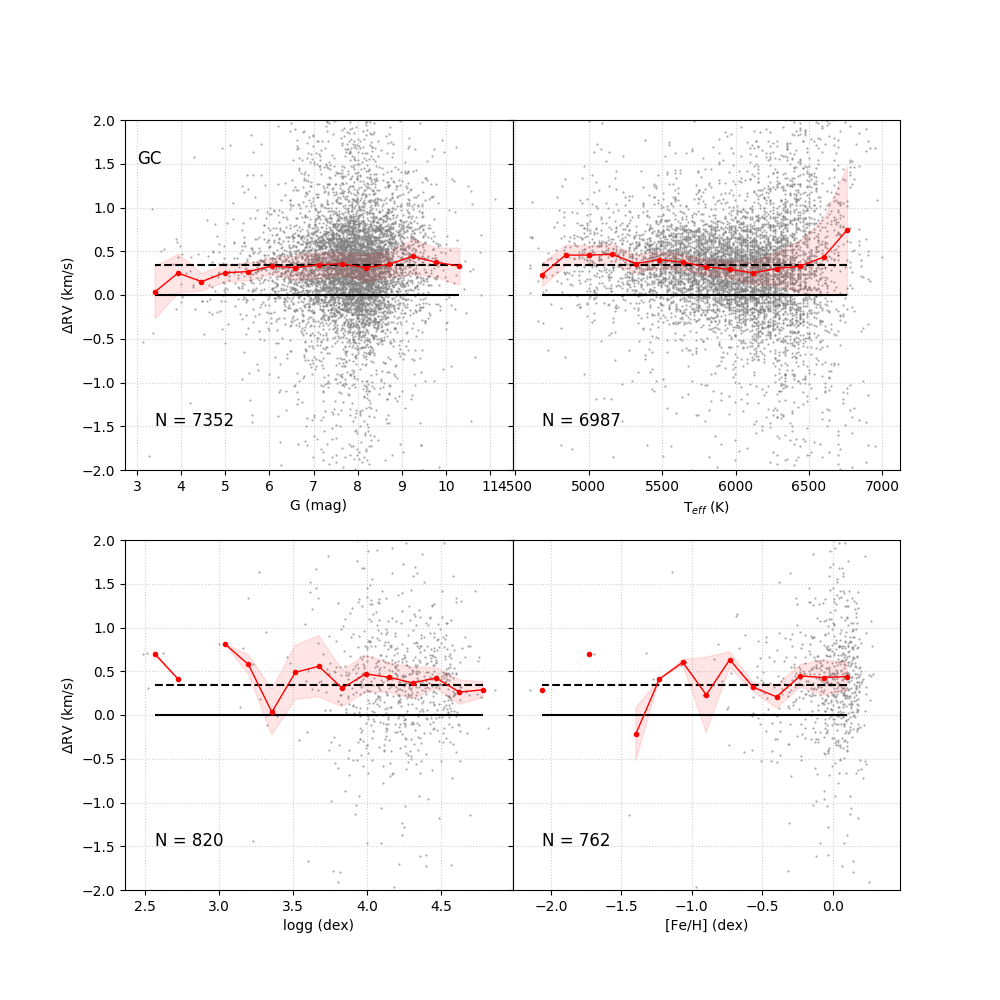}
\caption{RV comparison between SoS and the GC survey as a function of magnitude, $T_{\rm eff}$, $\log g$, and $[Fe/H]$. The red points are the median binned values and the shadowed area the MAD of each bin. The dashed line is their median difference.}
\label{GC}
\end{figure}

\begin{table}
\caption{Statistics for the RV differences of the external catalogues with SoS. N is the number of stars in common with SoS using the filter: flag\_rv$\neq$2.} 
\centering
\scalebox{0.95}{
\begin{tabular}{l c c c c c c}
\hline\hline
\multirow{2}{*}{Survey} & Mean           & Median & $\sigma$       & MAD &  \multirow{2}{*}{N} \\
                        & (km\,s$^{-1}$) & (km\,s$^{-1}$) & (km\,s$^{-1}$) & (km\,s$^{-1}$) & \\
\hline
   Gaia\,STD  & 0.31 & 0.34 & 0.64 & 0.16 & 4\,344 \\
   GC         & 0.82 & 0.67 & 2.96 & 0.31 & 7\,352 \\
\hline
\end{tabular}}
\label{table:9}
\end{table}

We perform a set for comparisons with external catalogues to confirm the robustness of our results. First, we compare SoS with the {\em Gaia} RV standard stars \citep[{\em Gaia}-STD,][]{Soubiran2018} which are comprised of 4813 FGK-type stars by combining numerous individual measurements from five high resolution spectrographs (R$>$45\,000): ELODIE \citep{Baranne1996} and SOPHIE \citep{Perruchot2008} on the 1.93\,m telescope at Observatoire de Haute-Provence (OHP), CORALIE on the Euler telescope at La Silla Observatory, HARPS at the ESO La Silla 3.6\,m telescope \citep{Mayor2003}, and NARVAL \citep{Auriere2003} on the Télescope Bernard Lyot at Pic du Midi Observatory. The ZP of {\em Gaia}-STD was established using asteroids whose RVs were accurately calculated from the dynamics of the solar system with an accuracy at a level of a few m\,s$^{-1}$. Their ZP is set at 0.38 km\,s$^{-1}$ which is interestingly very similar to the offset we find in our comparison of {\em Gaia}-STD with SoS (0.34 km\,s$^{-1}$, see Table~\ref{table:9}). We define this value as our absolute ZP and shift our final RVs to be in agreement with the {\em Gaia}-STD. Figure~\ref{gaia_standards} shows the difference in RVs between the two samples as a function of the stellar atmospheric parameters and magnitude. The stellar parameters in this plot are average values taken from spectroscopic surveys. We notice a very good agreement in the RVs with $\sigma$ of 0.64\,km\,s$^{-1}$ and MAD of 0.16\,km\,s$^{-1}$. 

The Geneva-Copenhaghen (GC) survey provides, among other parameters, very good quality RVs for G and F type dwarf stars in the solar neighbourhood \citep{Nordstrom2004}. The catalogue contains 14\,139 RVs for disc stars obtained with the photoelectric cross-correlation spectrometers CORAVEL \citep{Baranne1979, Mayor1985} operated at the Swiss 1\,m telescope at OHP and at the Danish 1.5\,m telescope at La Silla in high resolution. The GC survey has 8097 stars in common with SoS but we used 7352 stars for the comparison (see Table~\ref{table:9}) by excluding the spectroscopic binaries flagged in both samples. In Fig.~\ref{GC}, we show the differences in RV as a function of magnitude, $T_{\rm eff}$, $\log g$, and $[Fe/H]$, respectively. The $T_{\rm eff}$ for this comparison is obtained from the GC catalogue because it covers most stars whereas $\log g$ and $[Fe/H]$ come from the spectroscopic surveys included in the SoS. The plots in Fig.~\ref{GC} appear flat even for the $\log g$ which shows a slight trend for dwarf stars in the comparison with {\em Gaia}-STD. Similar behaviour is presented for the $T_{\rm eff}$ comparisons of SoS with {\em Gaia}-STD and GC. The median RV difference of SoS with GC is 0.67 km\,s$^{-1}$ but after the ZP shift of the SoS it reaches 0.33\,km\,s$^{-1}$. We note that the differences between the {\em Gaia}-STD and GC is 0.32\,km\,s$^{-1}$ which is in perfect agreement with our results after correcting for the ZP defined by asteroids. 

The differences and the scatter in RVs for both comparisons at this level can be attributed to the limitations of spectroscopic measurements by gravitational shift, convective shift, other astrophysical effects (stellar rotation, stellar activity, granulation, etc.), low-mass companions (stars and exoplanets), and last but not least, in case of systematic offsets, to instrumental effects. The dispersion in $\Delta$RVs can be further decreased if we use the cleaner sample of SoS (flag\_rv$=$0).

\subsection{Future updates and the SEGUE survey}

\begin{figure}
\hspace{-0.4cm}
\includegraphics[width=10.2cm, keepaspectratio]{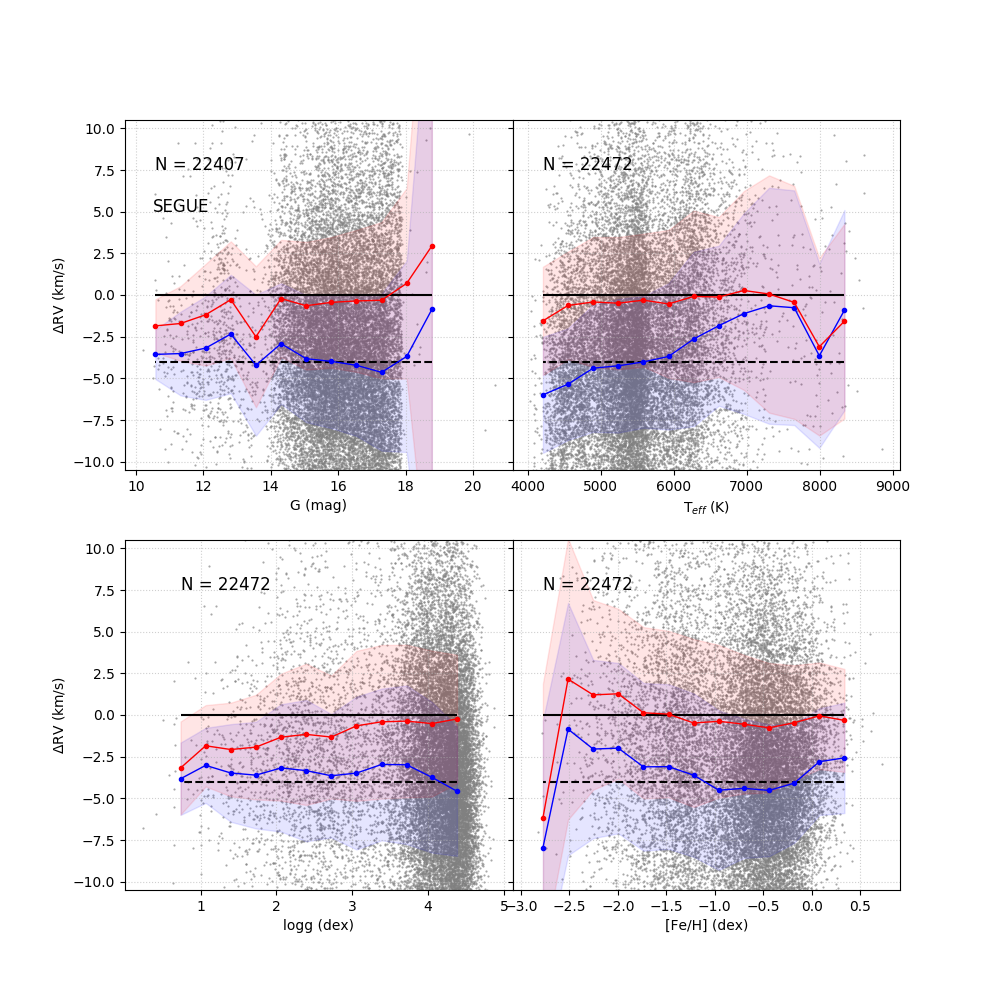} 
\caption{RV comparison between SoS and SEGUE as a function of magnitude, $T_{\rm eff}$, $\log g$, and $[Fe/H]$. The y-axis shows the $\Delta$RV = RV$_{SoS}$ - RV$_{SEGUE}$. The blue and red points represent the median $\Delta$RV of each bin with more than 3 entries before and after the calibration respectively. The blue and red shadowed areas are the MAD of each bin.}
\label{segue}
\end{figure}

This release of the SoS includes surveys with strong overlap with {\em Gaia}. However, there are other important spectroscopic surveys mapping the Galaxy with reliable parameters to be inserted in SoS in the future. For instance, the Sloan Extension for Galactic Understanding and Exploration \citep[SEGUE,][]{Yanny2009} and SEGUE-2 surveys have derived astrophysical parameters for $\sim$562\,000 stars in low resolution (R\,$\sim$\,2\,000). The overlap of SEGUE with {\em Gaia} stars with available RVs is small because SEGUE targets much fainter stars. This survey cannot serve well for the {\em Gaia} RV calibration methods we used previously but it can be inserted to SoS following a different procedure. In this case, the entire SoS is used as reference and any external catalogue is calibrated into the SoS reference frame. 

The stellar parameters for SEGUE are derived from the SEGUE Stellar Parameter Pipeline (SSPP)\footnote{Accessing SSPP Parameters: \url{http://www.sdss3.org/dr10/spectro/sspp\_data.php}} \citep{Lee2008a, Lee2008b, Prieto2008} with typical uncertainties in RVs of 2.4\,km\,s$^{-1}$. In Fig.~\ref{segue}, we show the RV comparisons of stars in common between SEGUE and SoS after flagging both samples for their high quality parameters. We find an zero point offset of --4.00\,km\,s$^{-1}$ with SoS and their differences mainly correlate with effective temperature and metallicity. Figure~\ref{segue} also shows the calibrations we apply to the SEGUE RVs in order to place them to the SoS RV frame after fitting linear functions as in Eq.~\ref{rvcoreq} (with $\alpha$\,=\,0 and $\zeta$\,=\,0). 

We are looking forward to upcoming surveys such as SDSS-V \citep{Kollmeier2017}, WEAVE \citep{Dalton2018}, 4MOST \citep{deJong2019}, and PFS \citep{Takada2014} which can be also appended to SoS. Depending on the size of the overlap with SoS and the accuracy of the above surveys, they can also be used to revise and update the current SoS reference system.   

\section{Science validation with open clusters}\label{science}

\begin{figure}
\centering
\includegraphics[width=9.0cm, clip]{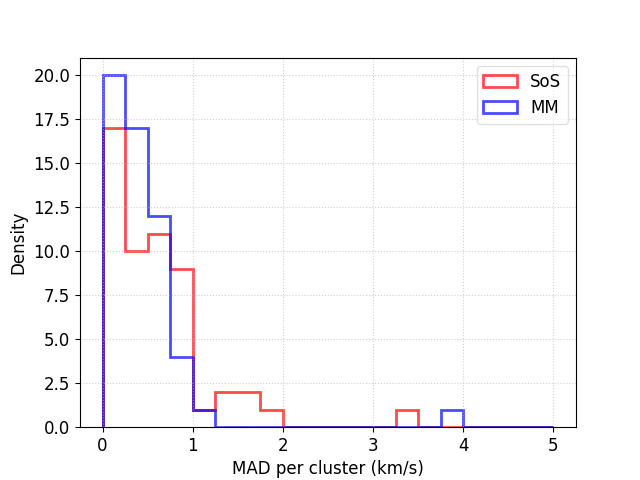}
\caption{The distribution of the MAD of the RVs derived from SoS in red and \cite{Mermilliod2008, Mermilliod2009} in blue for the 55 clusters. The OCs contain more than 3 stars and the MAD is calculated after a 3\,$\sigma$ outlier removal for both samples.}
\label{sigma_oc_mm}
\end{figure}

\begin{figure}
\centering
\includegraphics[width=10cm, clip]{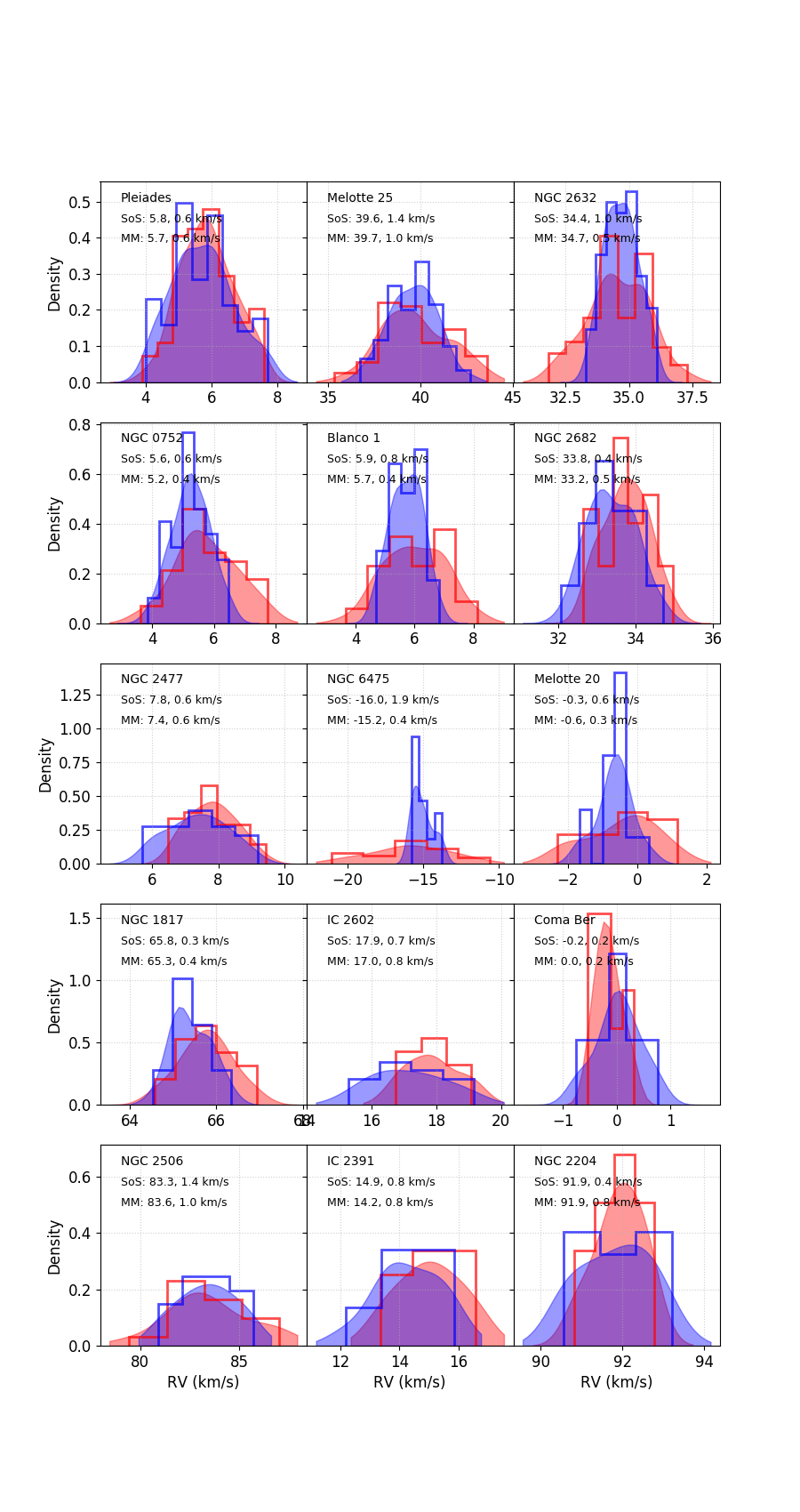}
\caption{RV distributions of 15 clusters from \cite{Mermilliod2008, Mermilliod2009} with more than 15 stars. Red histograms represent the SoS results and blue the literature. The Gaussian kernel-density estimate is plotted as shaded areas. There is also the information on the median and MAD values after the 3$\sigma$ outlier removal for both samples. }
\label{ocmm}
\end{figure}

\begin{figure}
\centering
\includegraphics[width=9.0cm, clip]{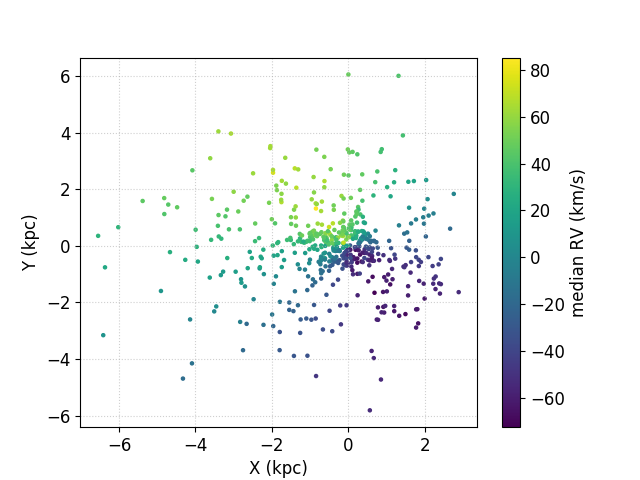} 
\caption{The spatial distribution of the 532 OCs observed with {\em Gaia} using the median RVs from SoS in color. The X and Y distances are taken from \cite{Cantat2020}. 
}
\label{ocgaia}
\end{figure}

\begin{table*}
\caption{The parameters of 532 OCs from \cite{Cantat2020} with more than 3 stars in SoS. N indicates the number of stars used for the calculation of the RV properties not the total number of members. The complete table is in electronic form.} 
\centering
\begin{tabular}{l r c c c c c}
\hline\hline
\multirow{2}{*}{Cluster} & RA & Dec  & RV$_{OC}$ & $\sigma$ & MAD & \multirow{2}{*}{N} \\
  & (deg)  & (deg)  & (km\,s$^{-1}$) & (km\,s$^{-1}$) & (km\,s$^{-1}$) & \\
\hline
ASCC\,10  & 51.562  & 34.962 & 34.93 $\pm$ 0.09 & 0.33 & 0.22 & 13 \\
ASCC\,101 & 288.370 & 36.343 & 36.44 $\pm$ 0.12 & 0.44 & 0.31 & 13 \\
ASCC\,105 & 296.038 & 27.630 & 27.59 $\pm$ 0.28 & 1.03 & 0.86 & 13 \\
ASCC\,11  & 53.055  & 44.939 & 44.88 $\pm$ 0.07 & 0.34 & 0.24 & 24 \\
ASCC\,113 & 318.096 & 38.750 & 38.72 $\pm$ 0.07 & 0.27 & 0.23 & 16 \\
ASCC\,12  & 72.544  & 41.665 & 41.68 $\pm$ 0.12 & 0.45 & 0.37 & 14 \\
   ...    & ...     & ...    & ...              & ...  & ...  & ... \\
\hline
\end{tabular}
\label{table:11}
\end{table*}

Open clusters (OCs) are good examples to showcase some of the applications of SoS since radial velocities along with proper motions allow the study of their three-dimensional kinematics, trace their orbits, and relate them to the spiral structure of the Galactic disc. Moreover, RVs have proved to be an efficient method for membership determination since the stars in OCs are formed together from the same material, sharing the same kinematics. \cite{Mermilliod2008, Mermilliod2009} present 1\,309 red giants in 166 OCs and 2\,565 solar-type dwarfs in 179 nearby OCs, respectively obtained from high resolution CORAVEL-type spectrographs at the OHP and at La Silla.

A comparison between homogeneous and independent literature analyses of OCs with SoS will demonstrate the precision and accuracy of our results, in particular when comparing the average RVs of the clusters and their dispersion. The sample of \cite{Mermilliod2008, Mermilliod2009}, hereafter MM, is ideal for this comparison because of its size and its high accuracy of 0.20-0.30\,km\,s$^{-1}$ \citep{Duquennoy1991, Baranne1996}. From this sample, we select clusters with more than 3 stars in common with SoS resulting in 1\,064 stars in 55 clusters, excluding stars flagged as non members and binaries by MM. We calculate their weighted mean RV (RV$_{OC}$) and MAD after a 3\,$\sigma$ outlier removal for both SoS and MM samples. The average difference of RV$_{OC}$ between SoS and MM is 0.21\,km\,s$^{-1}$ and the MAD of 0.26\,km\,s$^{-1}$. Figure~\ref{sigma_oc_mm} shows the distribution of the MAD for both SoS and MM for the 55 clusters to have very similar behaviour in both samples which indicates that they both are of similar accuracy and precision. As an example, we show the distribution of the RVs for the OCs with the highest number of stars in common (more than 15). The weighted mean RV and MAD as presented in Fig~\ref{ocmm} are very close for both samples which is a very good indication that our homogenization works.

We note that 27 from the 55 clusters have RVs in SoS derived from the homogenization of two or more surveys and the rest come from the homogenized {\em Gaia} RVs. The SoS RVs for the clusters in Fig.~\ref{ocmm} come from the homogenization of all six surveys apart from NGC 2506 and NGC 6475 where the RV source is only {\em Gaia}. Interestingly, these two clusters exhibit the highest MAD among the most numerous OCs in this sample. The results from Figs.~\ref{sigma_oc_mm}-\ref{ocmm} show that the homogenization included in SoS agrees within 0.26\,km\,s$^{-1}$ with the high quality CORAVEL studies, designed to reach an accuracy of 0.20-0.30\,km\,s$^{-1}$. Therefore, our homogenization and calibration procedures allowed us to effectively overcome the limitation of each individual survey.

The field of open clusters studies has received a tremendous boost thanks to {\em Gaia} data  using astrometric and kinematic criteria \citep[e.g.][]{Cantat2018, Liu2019, Castro2020, Cantat2020} providing catalogues with more than 2\,000 different clusters with hundreds of them newly discovered and have yet to be characterised \citep[e.g.][]{Soubiran2018b}. SoS in this release can significant contribute to this effort by investigating their kinematics for instance, by exploring a sample of a few thousands OCs from {\em Gaia} data with an accuracy of 0.26\,km\,s$^{-1}$ in the RV$_{OC}$. We select the recent sample of \cite{Cantat2020} comprising of 2\,014 OCs with member probability higher than 70\%. From this sample, we select clusters with more than 3 stars in common with SoS amounting to a total of 532 OCs. We note that not all OCs observed with {\em Gaia} have RVs measurements in {\em Gaia} DR2. Even though there may not be a full RV coverage by {\em Gaia} for these OCs, we incorporate in SoS ground based RVs which proves the synergy and the full exploitation of the available data. 

As in the previous OC analysis, we calculate the weighted mean and their corresponding errors in RVs per cluster shown in Table~\ref{table:11} and their distribution in the Galaxy in Fig.~\ref{ocgaia} also showing the rotation in the plane. Among these OCs there are some poorly studied such as some of the 41 new stellar clusters detected from \cite{Cantat2019} and some of the few hundreds from \cite{Castro2020}. 

This test is a first step in the study of OCs with SoS in a self-consistent way. In the next release we will provide homogeneous metallicities and abundances for these clusters. In fact, we already have iron metallicity measurements for 294 OCs from this sample making this a unique sample to trace the history of our Galaxy. 

\section{Data availability}

The SoS catalogue of RVs and the re-calibrated survey catalogues are available at  \url{http://gaiaportal.ssdc.asi.it/SoS}. 

\section{Conclusions}\label{conclusions}

SoS is a project to gather RV measurements from the largest spectroscopic surveys (APOGEE, GALAH, GES, RAVE, and LAMOST) including {\em Gaia} with the goal to deliver homogeneous determinations to the community in possibly the largest catalogue to date. Combining data from different catalogues is not trivial because different surveys suffer from different biases. An additional problem in our homogenization process was the large amount of data which forced us to find time- and resource-efficient solutions for the data manipulation. 
For our homogenization we followed the steps below: 
\begin{itemize}
    \item We pre-process the data sets to understand the strength and limitations of each survey and to select samples of reliable stars to use in the homogenization procedure.
    \item We perform the XM of the ground-based surveys with {\em Gaia} and evaluate its efficiency based on finding duplicated sources and assessing the best matches in {\em Gaia} based on magnitudes and RVs.
    \item We normalize the errors of the RVs of each survey based on \textit{i}) the repeated measurements and \textit{ii}) the TCH method. This process delivers normalization factors to multiply the RV errors of each survey to reveal that LAMOST and RAVE have overestimated their errors while APOGEE and GALAH have them underestimated, and {\em Gaia} and GES have better determined errors according to our analysis.  
    \item We first calibrate {\em Gaia} RVs with respect to magnitude, metallicity and effective temperature from the ground-based surveys separately by defining calibration coefficients to apply to the full {\em Gaia} DR2 sample. 
    \item Then we calibrate the RVs from the five ground-based surveys with respect to the new {\em Gaia} reference as a function of effective temperature, surface gravity, metallicity, and S/N using a multiple regression function. 
    \item The external ZP is set from the comparison to the {\em Gaia} RV standard stars. 
    \item We further validate the accuracy of SoS by comparing with the GC survey to be at 0.31\,km\,s$^{-1}$. Moreover, we calculated the median RVs of OCs from the MM sample and compare them and their MAD with SoS to find excellent agreement with this high resolution and homogeneous sample. We also provided median RVs and MAD for 532 OCs discovered by {\em Gaia}, some of them poorly studied.  
\end{itemize}

SoS contains RVs for around 11 million sources distributed in both hemispheres. The catalogue's precision is set by the error distribution which peaks at 0.09\,km\,s$^{-1}$ and its accuracy is set by comparison with external catalogues at 0.16-0.31\,km\,s$^{-1}$. Our next goal is to perform a similar analysis on the main stellar parameters for effective temperature, surface gravity, metallicity and chemical abundances to provide an unbiased compilation of the main properties of the stars in our Galaxy. 

\begin{acknowledgements}
We thank the anonymous referee for the useful comments and suggestions that helped improve this work. We thank Lucio Angelo Antonelli, Matteo Perri, Peter B. Stetson, Ricardo Carrera, Matteo Monelli, Germano Sacco, and Michele Fabrizio for their contributions in this work.  

This research has been partially supported by the following grants: MIUR Premiale "{\em Gaia}-ESO survey" (PI Sofia Randich), MIUR Premiale "MiTiC: Mining the Cosmos" (PI Bianca Garilli), the ASI-INAF contract 2014-049-R.O: "Realizzazione attivit\`a tecniche/scientifiche presso ASDC" (PI Angelo Antonelli), Fondazione Cassa di Risparmio di Firenze, progetto: "Know the star, know the planet" (PI Elena Pancino), and Progetto Main Stream INAF: "Chemo-dynamics of globular clusters: the {\em Gaia} revolution" (PI Elena Pancino). CG acknowledges support from the State Research Agency (AEI) of the Spanish Ministry of Science, Innovation and Universities (MCIU) and the European Regional Development Fund (FEDER) under grant AYA2017-89076-P. TM acknowledges financial support from the Spanish Ministry of Science and Innovation (MICINN) through the Spanish State Research Agency, under the Severo Ochoa Program 2020-2023 (CEX2019-000920-S).
\\

This work uses data from the European Space Agency (ESA) space mission {\em Gaia}. {\em Gaia} data are being processed by the {\em Gaia} Data Processing and Analysis Consortium (DPAC). Funding for the DPAC is provided by national institutions, in particular the institutions participating in the {\em Gaia} Multi-Lateral Agreement (MLA). We acknowledge the use of the public data products from RAVE (\url{https://www.rave-survey.org}). Funding for RAVE has been provided by: the Leibniz Institute for Astrophysics Potsdam; the Australian Astronomical Observatory; the Australian National University; the Australian Research Council; the French National Research Agency; the German Research Foundation (SPP 1177 and SFB 881); the European Research Council (ERC-StG 240271 Galactica); the Istituto Nazionale di Astrofisica at Padova; The Johns Hopkins University; the National Science Foundation of the USA (AST-0908326); the W. M. Keck foundation; the Macquarie University; the Netherlands Research School for Astronomy; the Natural Sciences and Engineering Research Council of Canada; the Slovenian Research Agency; the Swiss National Science Foundation; the Science \& Technology Facilities Council of the UK; Opticon; Strasbourg Observatory; and the Universities of Basel, Groningen, Heidelberg and Sydney.  

This work made use of GALAH data (\url{https://galah-survey.org} acquired through the Australian Astronomical Observatory, under programs: A/2013B/13 (The GALAH pilot survey); A/2014A/25, A/2015A/19, A2017A/18 (The GALAH survey). We acknowledge the traditional owners of the land on which the AAT stands, the Gamilaraay people, and pay our respects to elders past and present.

We acknowledge the use of the public data products from APOGEE (\url{https://www.sdss.org}). Funding for the Sloan Digital Sky Survey IV has been provided by the Alfred P. Sloan Foundation, the U.S. Department of Energy Office of Science, and the Participating Institutions. SDSS acknowledges support and resources from the Center for High-Performance Computing at the University of Utah. 

SDSS is managed by the Astrophysical Research Consortium for the Participating Institutions of the SDSS Collaboration including the Brazilian Participation Group, the Carnegie Institution for Science, Carnegie Mellon University, Center for Astrophysics | Harvard \& Smithsonian, the Chilean Participation Group, the French Participation Group, Instituto de Astrofísica de Canarias, The Johns Hopkins University, Kavli Institute for the Physics and Mathematics of the Universe / University of Tokyo, the Korean Participation Group, Lawrence Berkeley National Laboratory, Leibniz Institut für Astrophysik Potsdam, Max-Planck-Institut für Astronomie (MPIA Heidelberg), Max-Planck-Institut für Astrophysik (MPA Garching), Max-Planck-Institut für Extraterrestrische Physik, National Astronomical Observatories of China, New Mexico State University, New York University, University of Notre Dame, Observatório Nacional / MCTI, The Ohio State University, Pennsylvania State University, Shanghai Astronomical Observatory, United Kingdom Participation Group, Universidad Nacional Autónoma de México, University of Arizona, University of Colorado Boulder, University of Oxford, University of Portsmouth, University of Utah, University of Virginia, University of Washington, University of Wisconsin, Vanderbilt University, and Yale University.

Support to the development of the GES (\url{https://www.gaia-eso.eu}) has been provided in part by the European Science Foundation Gaia Research for European Astronomy Training (GREAT-ESF) Research Network Program.

Guoshoujing Telescope (the Large Sky Area Multi-Object Fiber Spectroscopic Telescope LAMOST) is a National Major Scientific Project built by the Chinese Academy of Sciences. Funding for the project has been provided by the National Development and Reform Commission. LAMOST \url{http://www.lamost.org}) is operated and managed by the National Astronomical Observatories, Chinese Academy of Sciences.   

This research has made use of the GaiaPortal catalogues access tool, Agenzia Spaziale Italiana (ASI) - Space Science Data Center (SSDC), Rome, Italy (\url{http://gaiaportal.ssdc.asi.it}). \\

In Sect.\ref{gaia_rv_correction} we used the Kapteyn packages \citep{KapteynPackage} for the fits. Some of the figures in this paper have been plotted using the healpy and HEALPix packages \citep[\url{https://healpy.readthedocs.io},][]{Zonca2019, Gorski2005}. We also used the python packages: astropy \citep[\url{http://www.astropy.org},][]{astropy2013, astropy2018}, numpy \citep[\url{https://numpy.org},][]{2020NumPy-Array}, scipy \citep[\url{https://www.scipy.org},][]{2020SciPy-NMeth}, pandas \citep[\url{https://pandas.pydata.org},][]{mckinney-proc-scipy-2010}, and matplotlib \citep[\url{https://matplotlib.org},][]{Hunter:2007}. \\

We used NASA Astrophysics Data System Bibliographic Services, the arXiv pre-print server operated by Cornell University, and the VizieR catalogue access tool, CDS, Strasbourg, France (DOI: 10.26093/cds/vizier). 

\end{acknowledgements}

\bibliography{bibliography} 

\appendix

\section{Comparisons of $[Fe/H]$ and $T_{\rm eff}$ between surveys}\label{metal}

\begin{figure}
\centering
\includegraphics[width=6.5cm,clip]{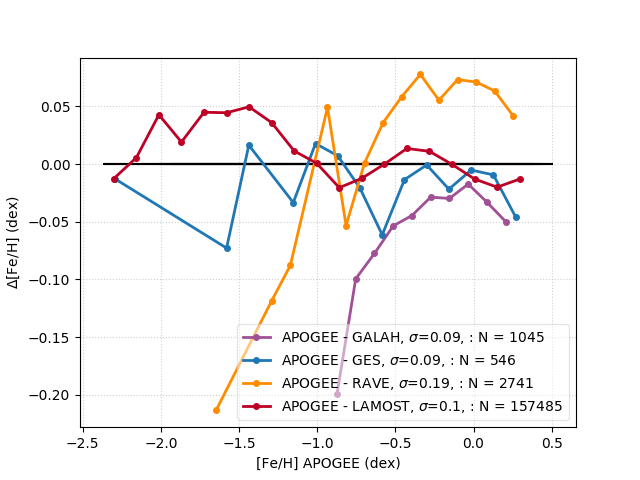} \\
\includegraphics[width=6.5cm,clip]{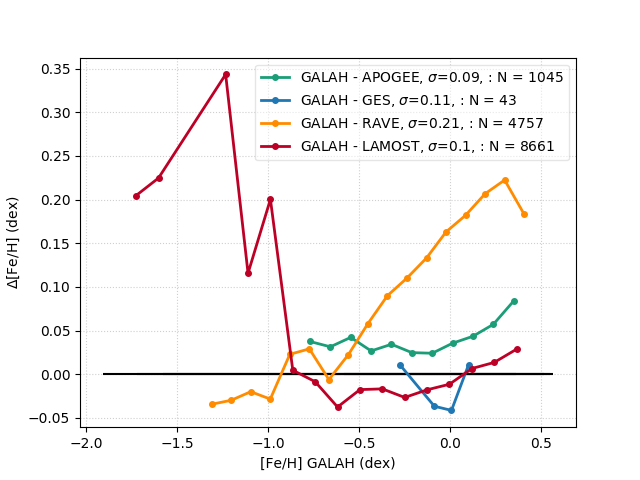} \\
\includegraphics[width=6.5cm,clip]{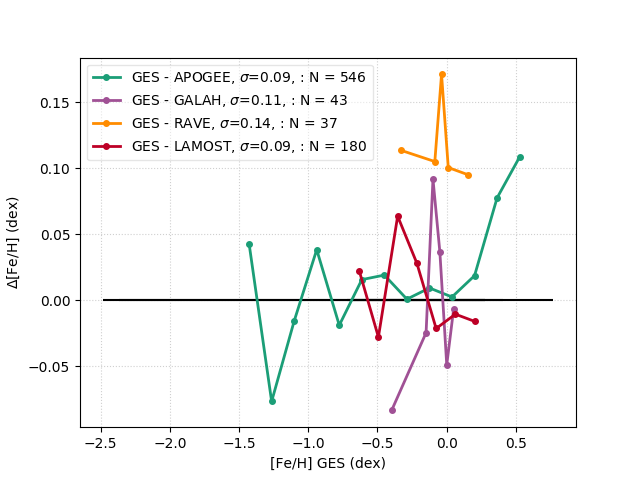} \\
\includegraphics[width=6.5cm, clip]{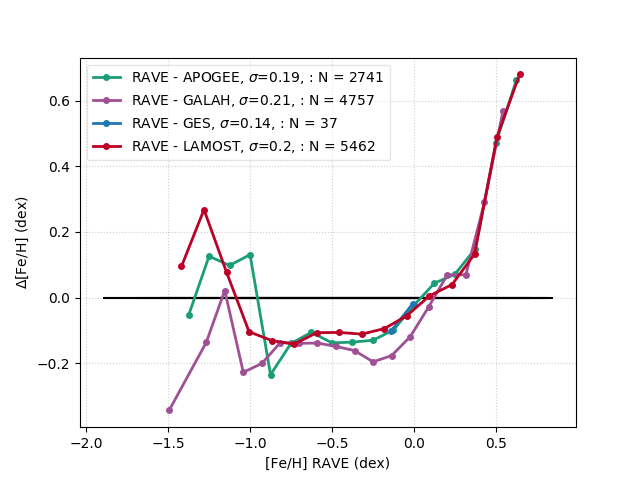} \\
\includegraphics[width=6.5cm,clip]{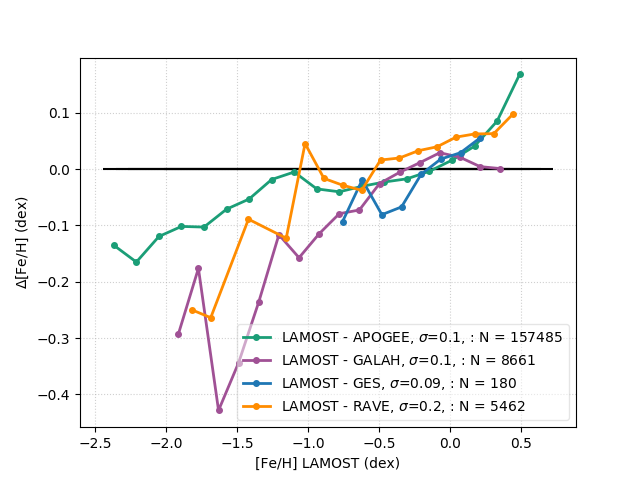} \\
\label{metal_rave}
\caption{Iron metallicity comparisons of stars in common for the paired surveys.}
\end{figure}

\begin{figure}
\centering
\includegraphics[width=6.5cm, clip]{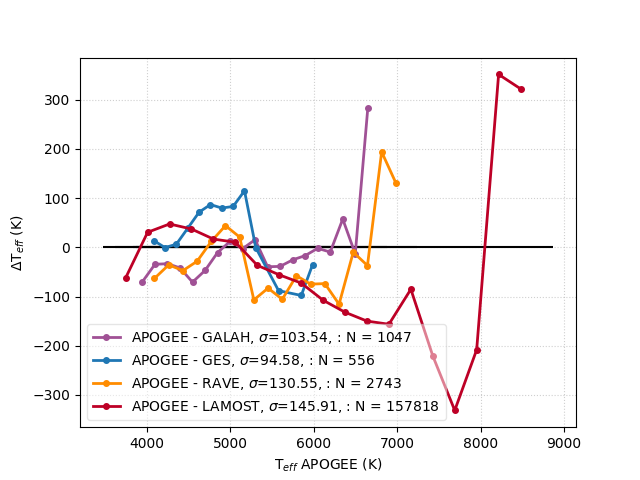}\\
\includegraphics[width=6.5cm, clip]{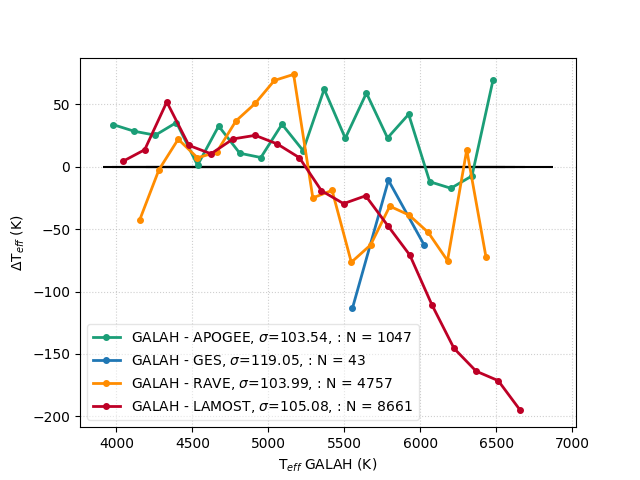} \\
\includegraphics[width=6.5cm, clip]{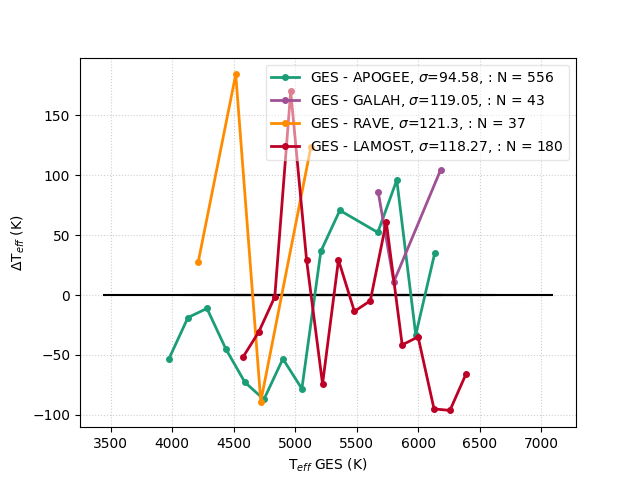}\\
\includegraphics[width=6.5cm, clip]{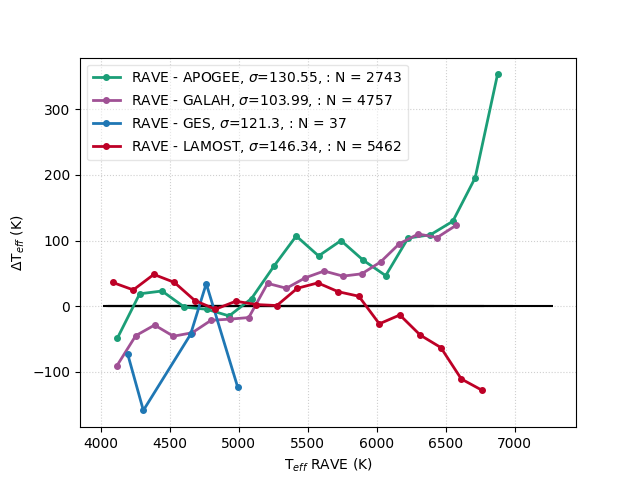} \\
\includegraphics[width=6.5cm, clip]{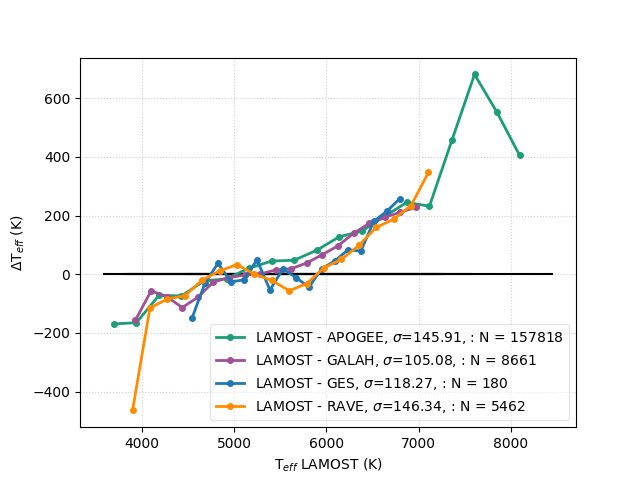} \\
\label{teff_comparison}
\caption{Differences in $T_{\rm eff}$ for stars in common for the paired surveys.}
\end{figure}

RAVE was primarily designed to be a Galactic archaeology survey focusing on obtaining reliable trends for populations of stars rather than providing precise stellar parameters for individual stars. 
Therefore, differences in the metallicity in comparison to the surveys are expected. In Fig.~\ref{metal_rave}, we present the differences in $[Fe/H]$ for the surveys in pairs. RAVE stars show the same trends in comparison to the other surveys. We apply a correction function for the calibration on the RAVE metallicities after fitting a second order polynomial:
\begin{equation}\label{rvgaiaeq}
    \Delta[Fe/H] = \alpha [Fe/H]_{RAVE}^2 + \beta [Fe/H]_{RAVE} + \gamma
\end{equation}
where $\alpha$ = --0.089 $\pm$ 0.002, $\beta$ = 0.432 $\pm$ 0.011, $\gamma$ = 0.55 $\pm$ 0.02

Similarly, the same comparisons are plotted for $T_{\rm eff}$ for the paired surveys in Fig.~\ref{teff_comparison}. In this case, we find LAMOST to show strong trends indicating that a correction of their $T_{\rm eff}$ scale is needed. As for RAVE above, we apply a linear fit to calibrate the LAMOST $T_{\rm eff}$:
\begin{equation}\label{lamost_teffeq}
    \Delta T_{\rm eff} = \alpha T_{\rm eff LAMOST} + \beta 
\end{equation}
where $\alpha$ = 0.122 $\pm$ 0.004, $\beta$ = --606 $\pm$ 2

\section{The RV calibration coefficients}\label{rvcalibrationappendix}

\begin{figure*}
\centering
\includegraphics[width=20cm, height=4.5cm, keepaspectratio]{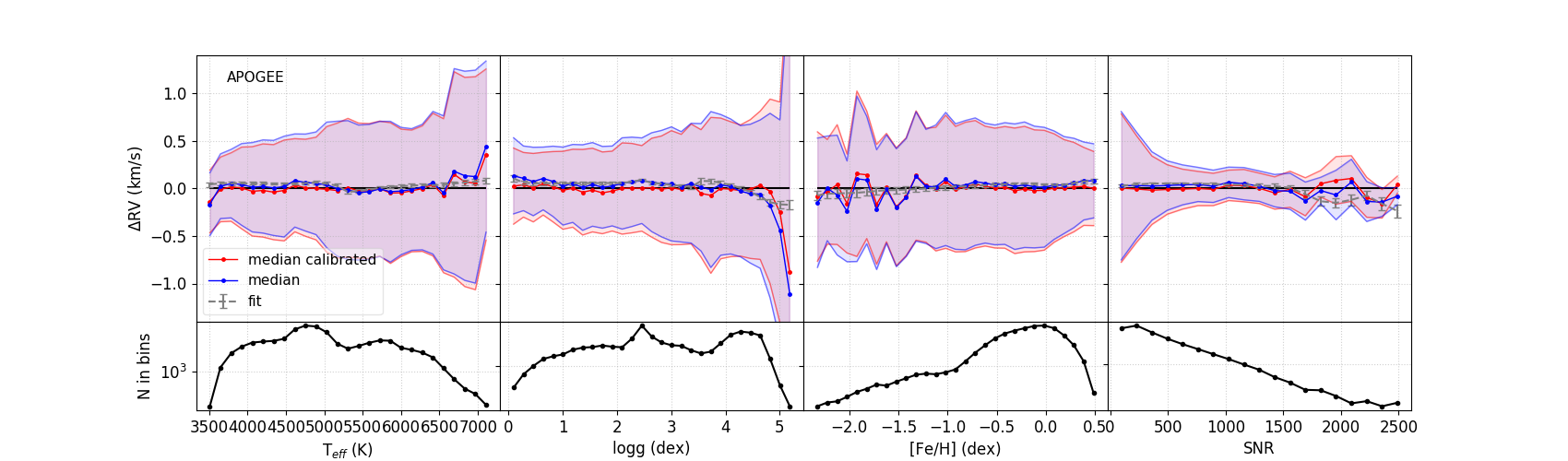} \\
\includegraphics[width=20cm, height=4.5cm, keepaspectratio]{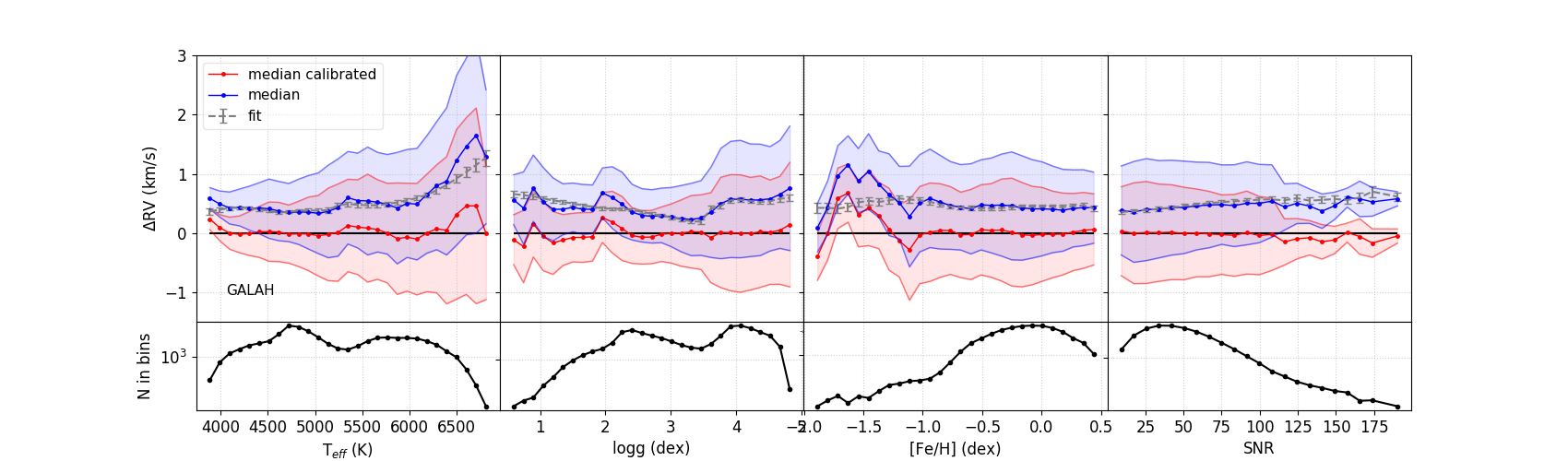}\\
\includegraphics[width=20cm, height=4.5cm, keepaspectratio]{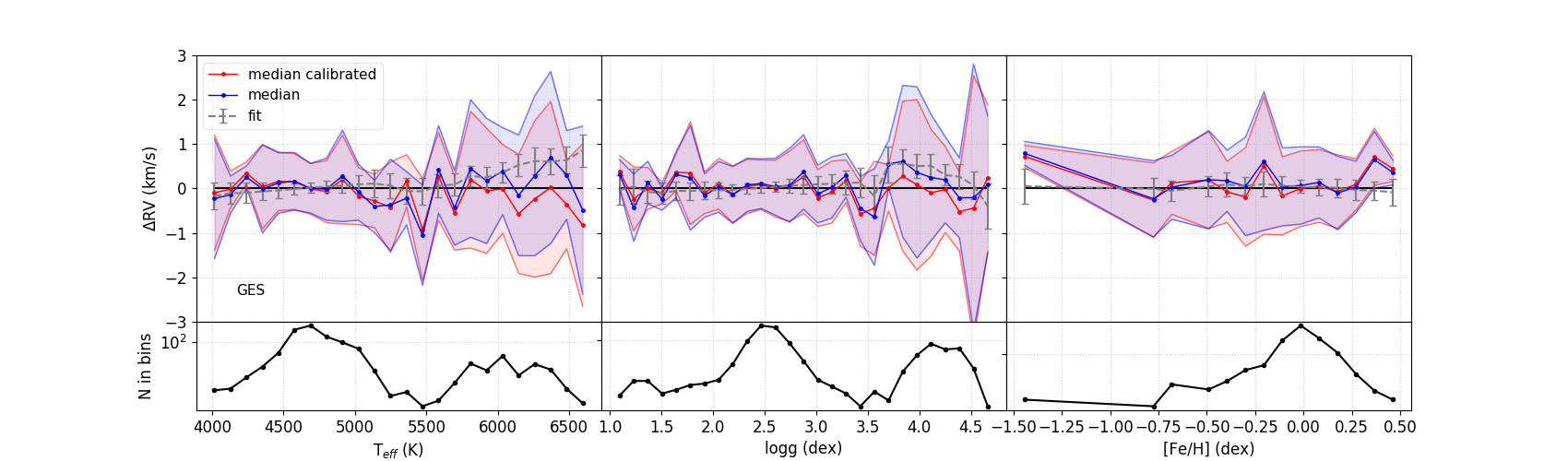} \\
\includegraphics[width=20cm, height=4.5cm, keepaspectratio]{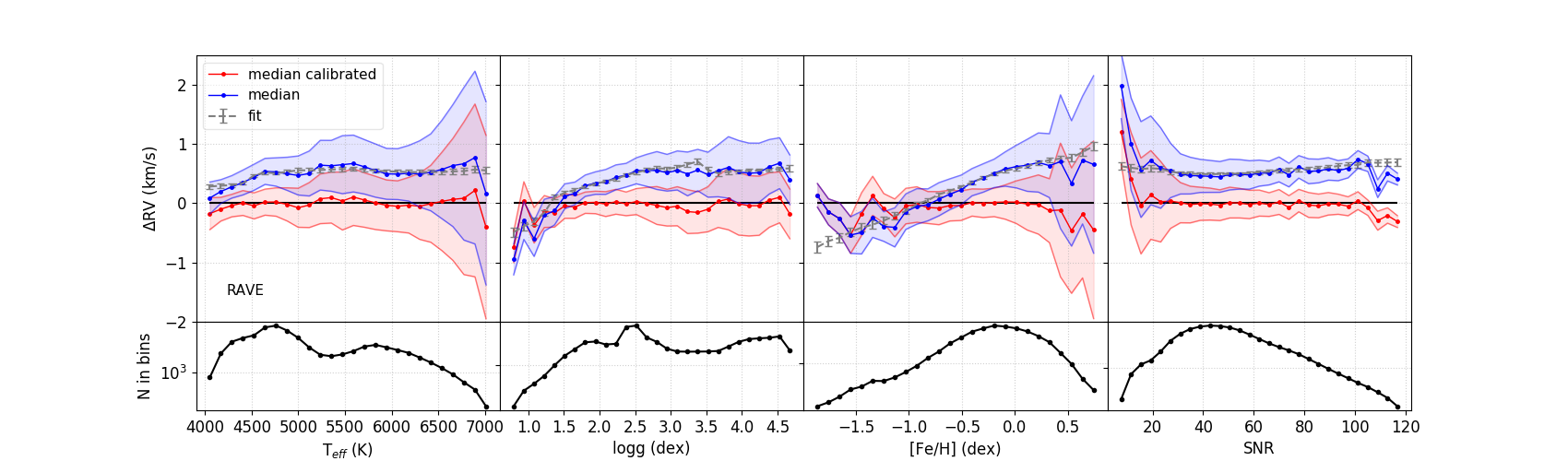}\\
\includegraphics[width=20cm, height=4.5cm, keepaspectratio]{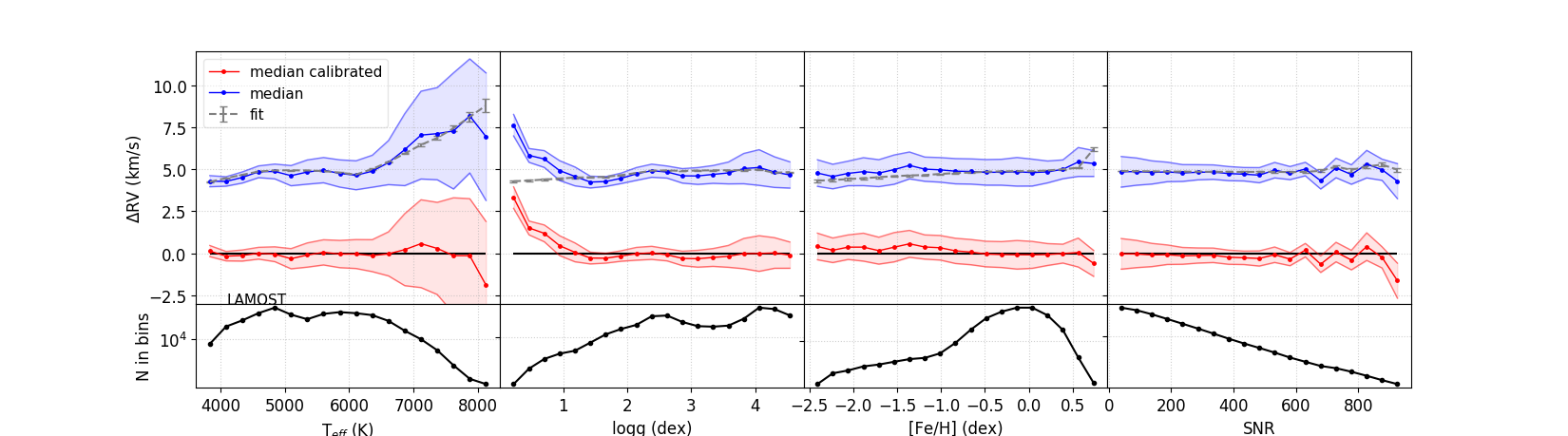}\\
\label{rvcor}
\caption{The calibration of the survey RVs for stars in common with {\em Gaia} based on Eq.~\ref{rvgaiateffeq} as a function of: $T_{\rm eff}$, $\log g$, $[Fe/H]$, and S/N. GES does not have S/N measurements. The $\Delta$RV in the y-axis indicates the \textit{calibrated} {\em Gaia} RVs minus the survey RVs before (blue points) and after the calibration (red points). The colors and symbols are also described in Fig.~\ref{rvgaia_mag}.}
\end{figure*}

\begin{figure}
\centering
\includegraphics[width=9.0cm,clip]{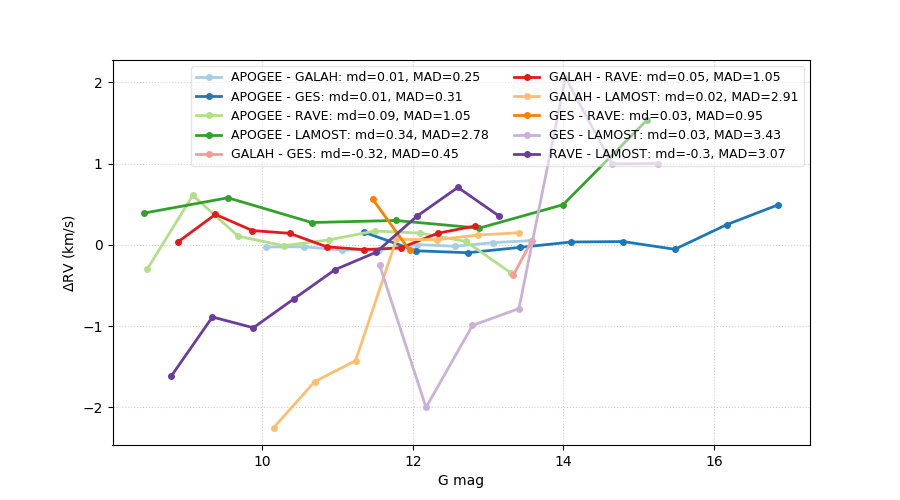} \\
\includegraphics[width=9.0cm,clip]{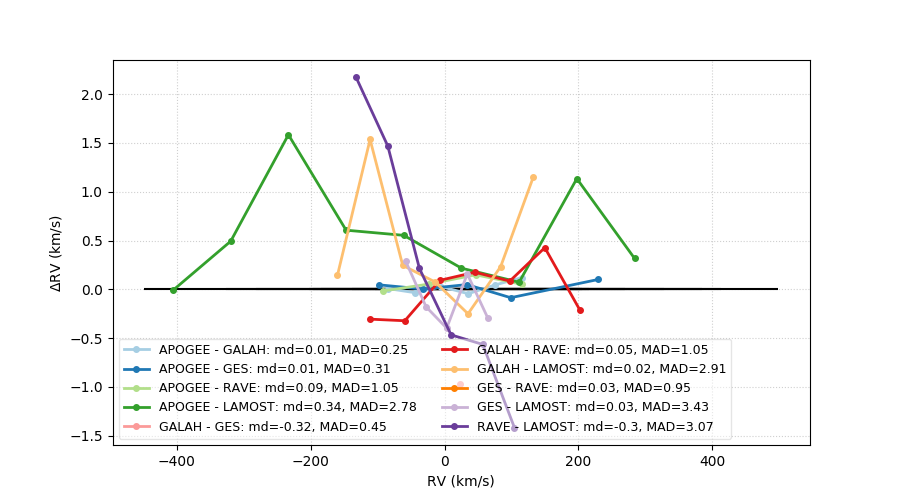} \\
\includegraphics[width=9.0cm,clip]{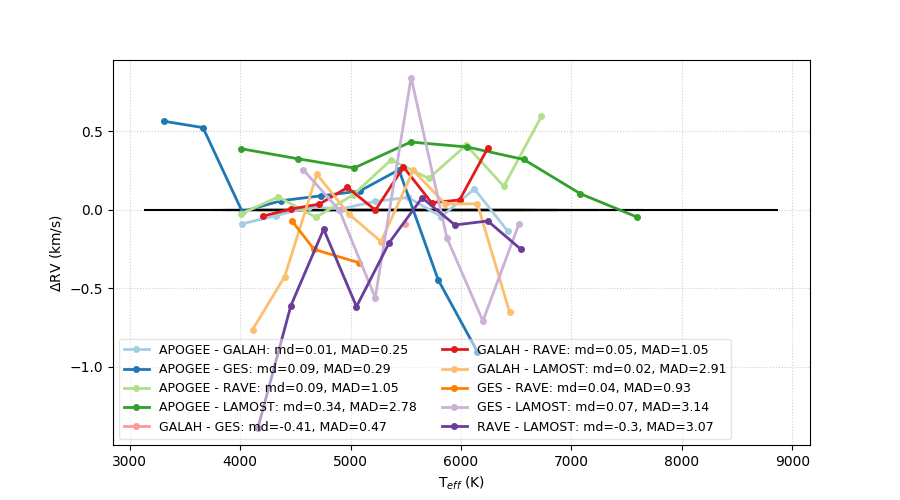} \\
\includegraphics[width=9.0cm,clip]{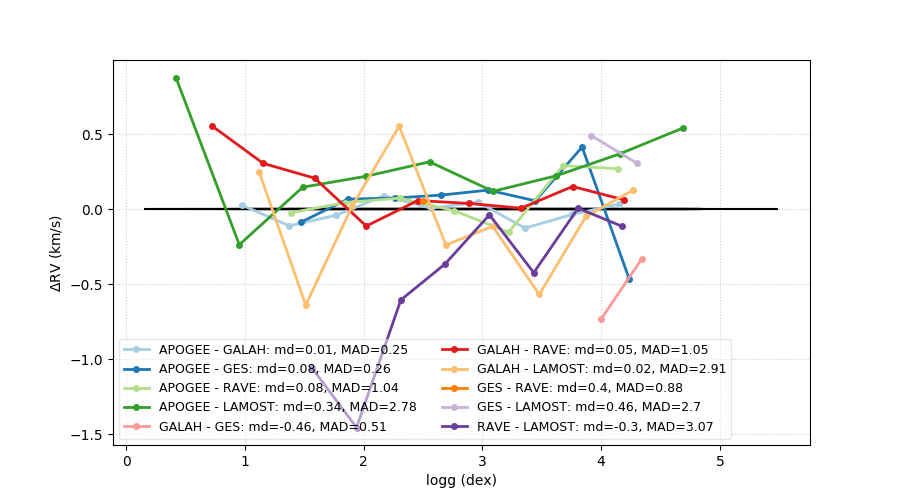} \\
\includegraphics[width=9.0cm,clip]{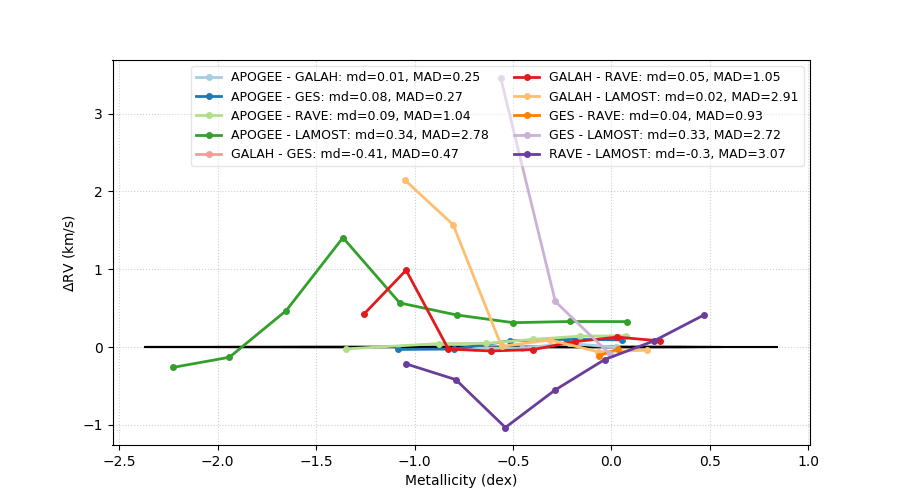} \\
\label{rvcor_pairedsurveys}
\caption{Comparison of the calibrated RVs for stars in common between surveys.}
\end{figure}

Table~\ref{table:5} summarises all the coefficients for the calibration of the {\em Gaia} RVs presented in Sect.~\ref{gaia_rv_correction} and the results of the fits are plotted in the respective Figs.~\ref{rvgaia_mag}-\ref{rvgaiacor_teff}. 
In turn, Table~\ref{table:8} shows the coefficients to calibrate the survey RVs  to the \textit{calibrated} {\em Gaia} RV reference frame, discussed in Sect.~\ref{surveys_rv_calibration}, and the results of the fits are plotted in Fig.~\ref{rvcor}. Additionally, we plot the calibrated RVs for the surveys for the stars in common in pairs in Fig~\ref{rvcor_pairedsurveys}. 

\begin{table}[h]
\caption{The coefficients for all the {\em Gaia} RV calibrations from Eqs.~\ref{rvmaggaiaeq}-\ref{rvgaiateffeq} in Sect.~\ref{gaia_rv_correction} for each survey and the final coefficients derived from their weighted mean.}
\centering
\scalebox{0.82}{
\begin{tabular}{l c c c c}
\hline\hline
\multicolumn{5}{c}{G\,mag calibration} \\
\hline
\multirow{2}{*}{Survey} & $\alpha$ & $\beta$ & $\gamma$ & \multirow{2}{*}{N} \\
 &  G mag${^2}$ & G mag & intercept & \\
\hline
   APOGEE & 0.497 $\pm$ 0.045 & --0.241 $\pm$ 0.009 & 0.0169 $\pm$ 0.0004 & 177\,407 \\
   GALAH  & 2.864 $\pm$ 0.350 & --0.680 $\pm$ 0.062 & 0.0364 $\pm$ 0.0028 & 91\,413 \\
   GES    & 1.055 $\pm$ 0.526 & --0.401 $\pm$ 0.114 & 0.0250 $\pm$ 0.0059 & 1\,208 \\
   RAVE   & 1.683 $\pm$ 0.134 & --0.403 $\pm$ 0.025 & 0.0222 $\pm$ 0.0012 & 436\,706 \\
   Final  & 0.638 $\pm$ 0.043 & --0.263 $\pm$ 0.008 & 0.0178 $\pm$ 0.0004 & - \\
\hline
\multicolumn{5}{c}{metallicity calibration} \\
\hline
       &    -    & $[Fe/H]$ & intercept & \\
\hline
APOGEE & - & --0.010 $\pm$ 0.002 & --0.140 $\pm$ 0.006 & 176\,863 \\
GALAH  & - & --0.044 $\pm$ 0.004 & --0.255 $\pm$ 0.012 & 87\,391 \\
GES    & - & --0.066 $\pm$ 0.035 & --0.222 $\pm$ 0.106 & 1\,211  \\
Final  & - & --0.018 $\pm$ 0.002 & --0.164 $\pm$ 0.006 & -  \\
\hline
\multicolumn{5}{c}{$T_{\rm eff}$ calibration} \\
\hline
       & $T_{\rm eff}^{2}$  & $T_{\rm eff}$ ($\times$10$^{-7}$) & intercept ($\times$10$^{-3}$) & \\
\hline
APOGEE & 4.21 $\pm$ 0.79 & --5.18 $\pm$ 1.01 & 15.9 $\pm$ 3.2 & 18\,982 \\
\hline
\end{tabular}}
\label{table:5}
\end{table}

\begin{table*}
\caption{The coefficients for the RV calibration of Eq.~\ref{rvcoreq} for each survey. The respective plots are shown in Fig.~\ref{rvcor}.}
\centering
\scalebox{0.86}{
\begin{tabular}{l c c c c c c r r }
\hline\hline
Survey & $\alpha$ $\times$10$^{-7}$ & $\beta$ $\times$10$^{-3}$ & $\gamma$ & $\delta$ & $\epsilon$ $\times$10$^{-4}$ & $\zeta$ $\times$10$^{-2}$ & $\eta$ & N \\
   
\hline
   APOGEE dwarfs & -                 &   0.040 $\pm$ 0.007 & --0.163 $\pm$ 0.014 & --0.082 $\pm$ 0.019 & --1.23 $\pm$ 0.13 & -                   &   0.49 $\pm$ 0.11 & 70\,220 \\
   APOGEE giants & -                 &   0.213 $\pm$ 0.017 & --0.134 $\pm$ 0.010 &   0.133 $\pm$ 0.009 & --0.40 $\pm$ 0.08 & -                   & --0.57 $\pm$ 0.06 & 104\,610 \\
   GALAH dwarfs  & 4.79   $\pm$ 0.37 & --5.217 $\pm$ 0.412 &   0.194 $\pm$ 0.024 &   0.132 $\pm$ 0.035 &   9.90 $\pm$ 2.88 & -                   &  13.78 $\pm$ 1.13 & 42\,868 \\
   GALAH giants  & --0.15 $\pm$ 0.44 &   0.288 $\pm$ 0.402 & --0.257 $\pm$ 0.015 &   0.108 $\pm$ 0.020 &  13.98 $\pm$ 1.60 & -                   & --0.06 $\pm$ 0.92 & 43\,961 \\
   GES dwarfs    & -                 &   0.607 $\pm$ 0.161 &   0.226 $\pm$ 0.313 & --1.419 $\pm$ 0.484 & -                 & -                   & --4.20 $\pm$ 1.77 & 379 \\
   GES giants    & -                 &   0.173 $\pm$ 0.329 &   0.061 $\pm$ 0.208 & --0.094 $\pm$ 0.166 & -                 & -                   & --0.95 $\pm$ 1.11 & 1\,195 \\
   RAVE dwarfs   & -                 & --0.054 $\pm$ 0.011 &   0.083 $\pm$ 0.017 &   0.549 $\pm$ 0.026 &  38.49 $\pm$ 3.81 & -                   &   0.32 $\pm$ 0.10 & 64\,906 \\
   RAVE giants   & -                 &   0.301 $\pm$ 0.023 &   0.012 $\pm$ 0.014 &   0.717 $\pm$ 0.013 &  20.96 $\pm$ 1.68 & -                   & --0.90 $\pm$ 0.08 & 12\,8063 \\
   LAMOST cool   & 4.41 $\pm$ 0.78   & --3.347 $\pm$ 1.040 & --1.451 $\pm$ 0.072 & --1.767 $\pm$ 0.045 & --3.99 $\pm$ 0.70 & --3.589 $\pm$ 0.030 &  14.16 $\pm$ 3.47 & 193\,357 \\
   LAMOST hot    & --5.24 $\pm$ 0.07 &   5.679 $\pm$ 0.075 & --0.147 $\pm$ 0.004 &   0.207 $\pm$ 0.010 & --7.50 $\pm$ 0.31 & --2.437 $\pm$ 0.068 & --9.88 $\pm$ 0.19 & 814\,681 \\
\hline
\end{tabular}}
\label{table:8}
\end{table*}

\section{Binarity parameters}\label{binaries_appendix}

\begin{figure}
\includegraphics[width=11cm, clip]{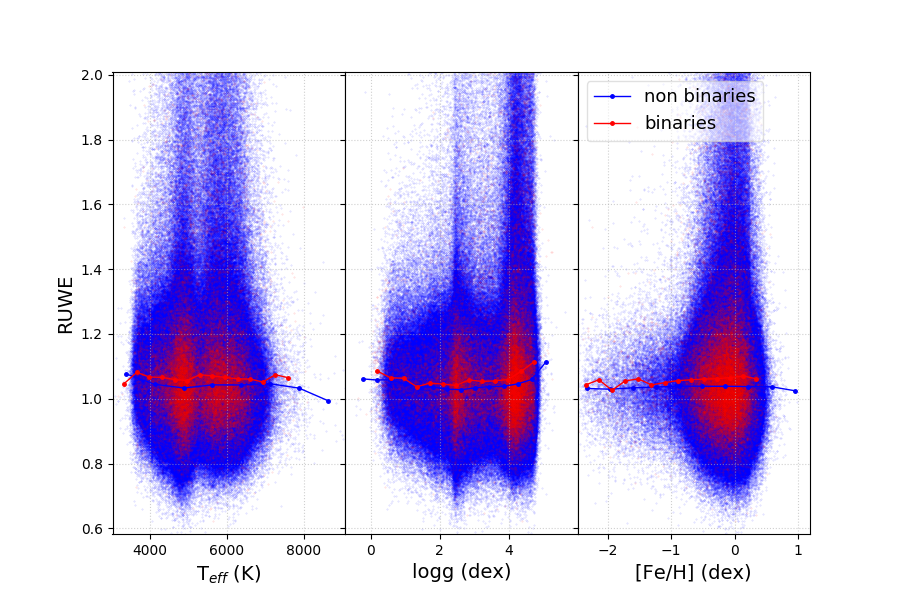} 
\caption{Binaries in SoS. The y axis is the RUWE for stars which have {\em Gaia} parameters. The x axis shows the stellar parameters from the surveys. The red points are the median binned values for the binary population and the blue points the rest of the SoS sample defined here as non binaries.}
\label{ruwe_sos}
\end{figure}

Another parameter which could infer binarity is the Renormalised Unit Weight Error (RUWE) which assesses the quality and reliability of the astrometric data calculated from the {\em Gaia} DR2. Binarity is one of the reasons for inconsistent astrometric solutions causing large RUWE values ($>$\,1.4). We have calculated the RUWE parameter for stars with {\em Gaia} parameters and find that in fact binary stars show higher RUWE (see Fig.~\ref{ruwe_sos}).

\end{document}